\documentclass{ieeeaccess}
\usepackage{cite}
\usepackage{amsmath,amssymb,amsfonts}
\usepackage{algorithm}
\usepackage{algorithmic}
\usepackage{graphicx}
\usepackage{textcomp}

\usepackage{lineno}
\usepackage{graphicx}
\usepackage{subfigure}
\usepackage{amssymb}
\usepackage{epstopdf}
\usepackage{booktabs}
\usepackage{caption}
\usepackage{wrapfig}
\usepackage{makecell}
\usepackage{CJK}
\usepackage{amsmath}
\usepackage{bm}
\usepackage{txfonts}
\usepackage{color}

\def\BibTeX{{\rm B\kern-.05em{\sc i\kern-.025em b}\kern-.08em
    T\kern-.1667em\lower.7ex\hbox{E}\kern-.125emX}}
\begin{document}
\history{Date of publication xxxx 00, 0000, date of current version xxxx 00, 0000.}
\doi{10.1109/ACCESS.2017.DOI}

\title{Open-Source Multi-Access Edge Computing for 6G: Opportunities and Challenges}
\author{\uppercase{Liqiang Zhao}\authorrefmark{1},
\IEEEmembership{Member, IEEE},
\uppercase{Guorong Zhou}\authorrefmark{1},
\uppercase{Gan Zheng}\authorrefmark{2},
\IEEEmembership{Fellow, IEEE},
\uppercase{Chih-Lin I}\authorrefmark{3},
\IEEEmembership{Fellow, IEEE},
\uppercase{Xiaohu You}\authorrefmark{4},
\IEEEmembership{Fellow, IEEE},
\uppercase{Lajos Hanzo}\authorrefmark{5},
\IEEEmembership{Fellow, IEEE}}

\address[1]{The State Key Laboratory of Integrated Services Networks, Xidian University, Xi'an 710071, China (e-mail: lqzhao@mail.xidian.edu.cn, guor\_zhou@163.com)}
\address[2]{The Wolfson School of Mechanical, Electrical and Manufacturing Engineering, Loughborough University, Loughborough LE11 3TU, U.K. (e-mail: g.zheng@lboro.ac.uk)}
\address[3]{Green Communication Research Center, China Mobile Research Institute, Beijing 100053, China (e-mail: icl@chinamobile.com)}
\address[4]{The National Mobile Communications Research Laboratory, Southeast University, Nanjing 210096, China (e-mail: xhyu@seu.edu.cn)}
\address[5]{The School of Electronics and Computer Science, University of Southampton, Southampton SO17 1BJ, U.K. (e-mail: lh@ecs.soton.ac.uk)}

\tfootnote{This work was supported in part by National Key R\&D Program of China (2019YFE0196400), Joint Project of China Mobile Research Institute \& X-NET (R202111101112JZC04), National Natural Science Foundation of China (61771358, 61901317, 62071352), Fundamental Research Funds for the Central Universities (JB190104), Science and Technology Plan of Xi'An City (2019217014GXRC006CG007-GXYD6.1), Joint Education Project between China and Central-Estern European Countries (202005) and the 111 Project (B08038).
L. Hanzo would like to acknowledge the financial support of the Engineering and Physical Sciences Research Council projects EP/P034284/1 and EP/P003990/1 (COALESCE) as well as of the European Research Council's Advanced Fellow Grant QuantCom (Grant No. 789028).}

\markboth
{L. Zhao \headeretal: Open-Source Multi-Access Edge Computing for 6G: Opportunities and Challenges}
{L. Zhao \headeretal: Open-Source Multi-Access Edge Computing for 6G: Opportunities and Challenges}

\corresp{Corresponding author: Lajos Hanzo (e-mail: lh@ecs.soton.ac.uk).}

\begin{abstract}
Multi-access edge computing (MEC) is capable of meeting the challenging requirements of next-generation networks, \textit{e.g.}, 6G, as a benefit of providing computing and caching capabilities in the close proximity of the users. However, the traditional MEC architecture relies on specialized hardware and its bespoke software functions are closely integrated with the hardware, hence it is too rigid for supporting the rapidly evolving scenarios in the face of the demanding requirements of 6G.
As a remedy, we conceive the compelling concept of open-source cellular networking and intrinsically amalgamate it with MEC, which is defined by open-source software running on general-purpose hardware platforms.
Specifically, an open-source MEC (OS-MEC) scheme is presented relying on a pair of core principles:
the decoupling of the MEC functions and resources from each other with the aid of network function virtualization (NFV);
as well as the reconfiguration of the disaggregated MEC functions and resources into customized edge instances.
This philosophy allows operators to adaptively customize their users' networks.
Then, we develop improved networking functions for OS-MEC decoupling and discuss both its key components as well as the process of OS-MEC reconfiguration.
The typical use cases of the proposed OS-MEC scheme are characterized with the aid of a small-scale test network.
Finally, we discuss some of the potential open-source-related technical challenges when facing 6G.
\end{abstract}


\titlepgskip=-15pt

\maketitle

\section{Introduction}
\label{sec:introduction}
\PARstart{B}{y} integrating communication and information technologies, multi-access edge computing (MEC) \cite{1a,111,222} intrinsically amalgamates the computing and caching capabilities in the immediate proximity of the users, which reduces the transmission delay and improves the resource utilization at the edge.
As a further benefit, the emergence of MEC alleviates both the potential data transmission congestion of the transport network (TN) as well as the data processing burden imposed on the core network (CN).
Therefore, in December 2014, the European telecommunications standards institute (ETSI) has initiated MEC standardization for continuously updating and improving the MEC reference architecture and its applications \cite{1b}.
Many operators and vendors provide a convenient computing and caching environment at the edge by relying on ETSI's MEC architecture \cite{21,22,23}.

With the continual growth of the mobile Internet traffic, next-generation networks are expected to operate at rates up to Tbps and delays lower than 1ms \cite{1,2,30}.
MEC is considered as a potential enabling technology of the sixth generation (6G), capable of satisfying enhanced service requirements \cite{333}.
For example, MEC can guarantee ultra-low transmission delay and satisfy flawless 4k/8k video broadcast by processing computation-intensive data or caching ultra-high-definition (UHD) videos at the edge \cite{24,25}.
However, the exiting MEC reference architecture is built upon specialized hardware and its software functions are closely integrated with the hardware, which is inconvenient in the face of rapidly evolving scenarios, when aiming for satisfying the demanding requirements of 6G.
As a remedy, the above challenges may be met by introducing the open-source (OS) cellular network concept into MEC for allowing operators to adaptively customize their users' networks.

Open-source software \cite{26} has a very long history, which can be traced back to the organizations of SHARE and DECUS in 1950s.
In the following years, ARPANET, UNIX, Linux and Android were developed for altruistic public sharing.
By contrast, OS cellular networks are still in their infancy, despite the associated benefits.
Explicitly, in the OS cellular network, an open infrastructure would enable flexible network reconfiguration based on the specific requirements of diverse scenarios, and open source software would enable users to develop the future network in a collaborative open-access manner.
Such an OS cellular network has to rely on the integration of 6G candidate technologies, such as network function virtualization (NFV) \cite{3}, software-defined networking (SDN) \cite{4}, network slicing \cite{4a} and MEC \cite{6}, \textit{etc.}
Among them, SDN/NFV serve as the core drivers, which should additionally be integrated with MEC for cell-edge performance enhancement in support of customized radio access networks (RANs).
\newcommand{\tabincell}[2]{\begin{tabular}{@{}#1@{}}#2\end{tabular}}
\begin{table*}[!tbp]
\centering
\caption*{TABLE 1: A comparative summary of contributions of the salient existing works.}
\begin{tabular}{|l|c|c|c|c|c|c|c|c|c|c|}
  \hline
  Feature&\tabincell{c}{OAI\\ \cite{11}}&\tabincell{c}{Open5G\\ \cite{12}}&\tabincell{c}{O-RAN \\ \cite{7aa,13}}& \tabincell{c}{ONAP \\ \cite{14}} & \tabincell{c}{OPNFV \\ \cite{15}}& \tabincell{c}{ETSI's \\ MEC\cite{1b}} & \tabincell{c}{Akraino \\ \cite{6d}}& \tabincell{c}{SDEC \\ \cite{18}}& \tabincell{c}{NDN- \\ECC\cite{19}} & Proposed \\
  \hline
  Focusing on MEC &  &  &  &  &  & \checkmark & \checkmark & \checkmark & \checkmark & \checkmark \\
  \hline
  \tabincell{l}{Applied in 6G network} &  &  & \checkmark & \checkmark & \checkmark &  & \checkmark &  & \checkmark & \checkmark \\
  \hline
  \tabincell{l}{Open-source architecture} & \checkmark & \checkmark & \checkmark & \checkmark & \checkmark &  & \checkmark &\checkmark & \checkmark & \checkmark \\
  \hline
  \tabincell{l}{Application and resource \\decoupling using NFV} & \checkmark & \checkmark &  & \checkmark & \checkmark & \checkmark & \checkmark & \checkmark &  & \checkmark \\
  \hline
  \tabincell{l}{Network functions \\decoupling by SBA} &  &  &  & \checkmark &  &  &  &  &  & \checkmark \\
  \hline
  \tabincell{l}{Using MANO technology} &  & \checkmark & \checkmark & \checkmark &  & \checkmark & \checkmark & \checkmark &  & \checkmark \\
  \hline
  \tabincell{l}{Reconfiguration by \\templates and instances} &  &  &  &  &  &  &  &  &  & \checkmark \\
  \hline
  \tabincell{l}{Supporting multiple \\access on RAN side} &  & \checkmark &  &  &  & \checkmark &  & \checkmark &  & \checkmark \\
  \hline
  \tabincell{l}{Focusing on upper \\layers of architecture} &  &  & \checkmark & \checkmark &  &  &  &  &  & \checkmark \\
  \hline
  \tabincell{l}{Using Docker and \\Kubernetes technologies} &  &  &  & \checkmark &  &  & \checkmark &  &  & \checkmark \\
  \hline
\end{tabular}
\end{table*}

Substantial efforts have been invested in promoting the progress of OS networks \cite{7}.
The Open-Air-Interface (OAI) community founded in Europe exploited a fully-fledged fourth generation (4G) protocol stack based on commercial off-the-shelf (COTS) hardware.
As a further advance, the Open5G community founded in China developed the concepts, technologies and platforms of open-source fifth generation (5G).
The O-RAN alliance \cite{7aa} founded by the global industry supported by academics has endeavoured to evolve RANs around the world.
Additionally, the Linux Foundation (LF) launched the open network automation platform (ONAP) for orchestration.
As for virtualization, the open platform of NFV (OPNFV) was conceived.
Recently, there have also been a few open-source efforts on MEC \cite{6d}.
For example, the Akraino project built by LF offers a pair of blueprints, namely the 5G MEC/slice and the micro-MEC system, but unfortunately it remains focused on the virtualization layer of the MEC reference architecture.
Based on all these efforts, we present a new paradigm for developing an all-encompassing open-source MEC (OS-MEC) scheme.

The contributions of this paper are summarized as follows.
\begin{enumerate}
\item Using ETSI's MEC architecture as a springboard, this paper introduces an OS-MEC framework to facilitate customized services for any potential scenario in 6G network, which is built on a pair of key principles, namely on network decoupling and reconfiguration.
\item With reference to the service-based architecture (SBA) of the 5G core network (5GC) \cite{6c}, a service-based MEC layer is developed for the OS-MEC scheme for decomposing the tightly coupled service functions into multiple independent network functions (NFs), thereby achieving MEC decoupling.
\item In order to integrate these decoupled NFs as required and to reconfigure a MEC system, the concept of templates and instances is implemented as part of the NF's management and orchestration (MANO) scheme, eventually generating a customized entity for supporting complete services.
\item Several typical use cases are demonstrated based on our test network for validating its flexibility and customization. Our novel contributions are also contrasted at a glance boldly and explicitly to the state-of-the-art in Table 1.
\end{enumerate}

The rest of this paper is organized as follows. We first introduce the basic concepts and key technologies of OS cellular networks in Section II. In Section III, we present our OS-MEC framework, its service-based MEC layer, interface, and the NF's MANO. In Section IV, our small-scale test network, our use cases and our demonstrations are presented.
In Section V, we elaborate on some potential technical challenges.
Finally, we conclude in Section VI.

\section{The Open-Source Network}

\begin{figure}[!tp]
\setlength{\abovecaptionskip}{0cm}
\setlength{\belowcaptionskip}{0cm}
  \centering
  \includegraphics[width=3.4in]{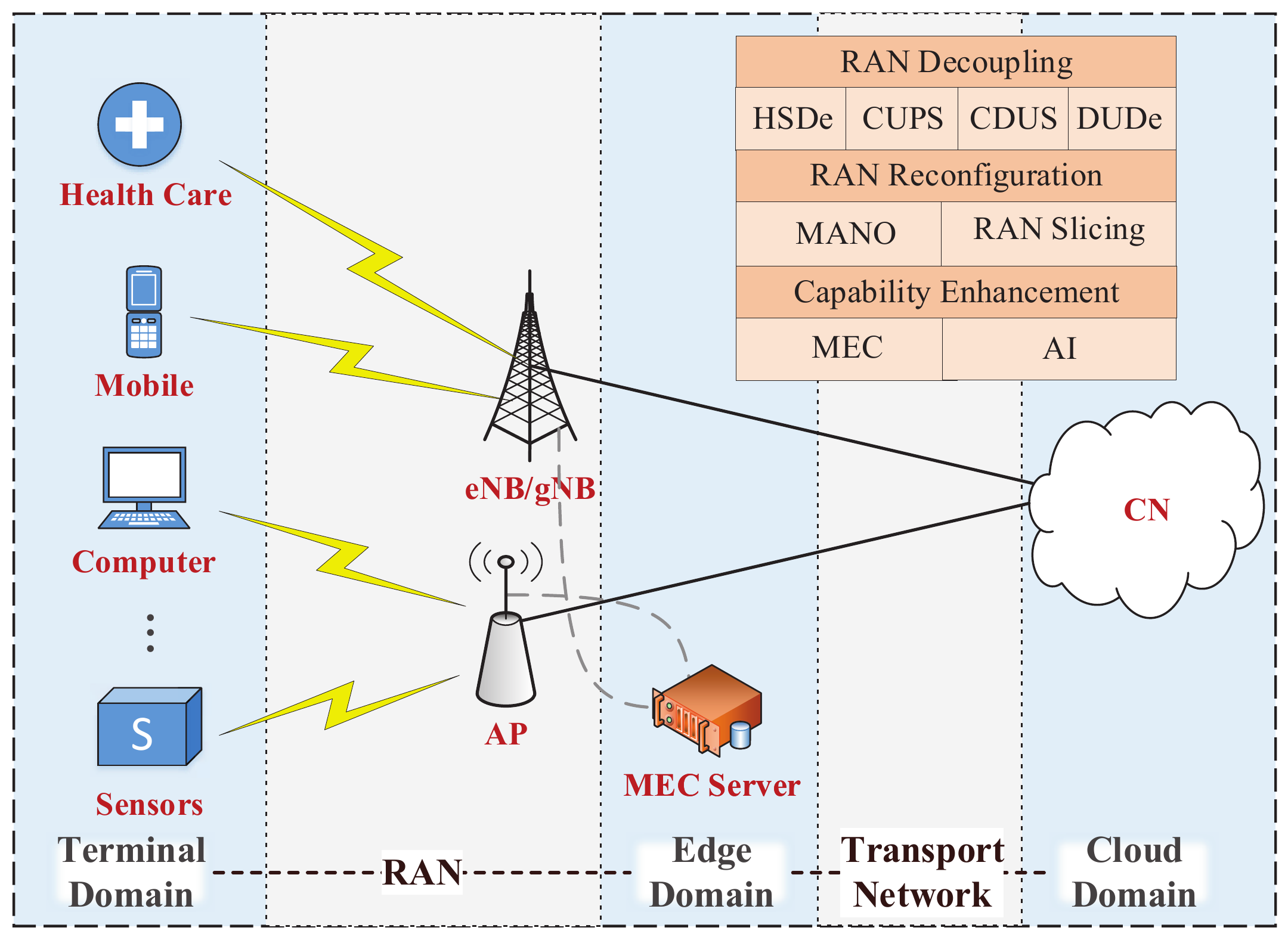}\\
  \caption{OS cellular network.}\label{fig1}
\end{figure}

In this section, our vision of an OS cellular network relies on four key elements:
the cloud domain including CN; the edge domain including distributed MEC servers; the heterogeneous terminal domain; and the network domain connecting the cloud-edge-terminal domain, which includes a RAN linking the edge and terminal domains, and a transport network linking the cloud and edge domains,
as shown in Fig.~\ref{fig1}.

In the cloud domain, the OS CN architecture has been well documented.
For example, SBA has been identified as a basic element of the 5GC architecture in the 3rd Generation Partnership Project's (3GPP) standards, which exhibits a high grade of flexibility and scalability.
In SBA, the originally coupled network functions are partitioned into multiple independent functional blocks, where each of them provides one or more services that can be used by any of the other functional blocks.
Therefore, all functional blocks can communicate with each other in order to support the users with customized applications, as if they were connected to a service bus.
In the edge domain, the OS-MEC concept will be discussed in the next section.
However, in the network domain, the OS RAN architecture has not as yet been defined, because it is hampered by its increased real-time complexity requirement.
To achieve this, some promising solutions, such as RAN decoupling and reconfiguration, have been put forward.
Firstly, the tightly coupled RAN elements, functions and resources of conventional networks have to be completely split.
Then, they could be reassembled dynamically according to a user's requirements, thus providing customized virtual RANs.
In a nutshell, the OS concept facilitates a convenient LEGO-block based construction of 6G RANs.

\subsection{RAN Decoupling}

The tightly coupled 4G cellular networks rely on a rigid closed operating system.
Hence at least four different RAN decoupling methods have been conceived for orchestrating a beneficial degree of openness in 6G, namely hardware/software decoupling (HSDe) \cite{5}, control plane and user plane splitting (CUPS) \cite{6b}, central unit and distributed unit splitting (CDUS) \cite{5b}, as well as downlink and uplink decoupling (DUDe) \cite{5c}.
Since they have different pros and cons, there is no wide agreement concerning the best OS RAN architecture.
It is worth noting that HSDe and CUPS constitute the basis of OS cellular networks, while CDUS and DUDe further exploit the flexibility and openness of cellular networks.

HSDe decouples the resources and functions from the physical facilities.
Given the mature family of NFV technologies, the network resources can be abstracted and shared among different users, and diverse services can be supported by the same infrastructure, thus separating the functionalities from the underlying infrastructure.

The idea of CUPS stems from SDN, which decouples the control signaling from the data transmission.
Then, it further expands to cellular networks by deploying control base stations (CBSs) in the most friendly lower frequency band and data base stations (DBSs) in higher frequency bands to form the control plane (CP) and user plane (UP), respectively.

CDUS represents one of the key technologies for the next-generation cloud RAN, which focuses on separating the traditional evolved node base station (eNB) into the radio transceiver (distributed unit) and the logical center (central unit) from the perspective of the OS protocol stack.

The tightly coupled downlink (DL) and uplink (UL) transmission limits the flexibility of a user's connections.
Upon adopting DUDe via dual-connectivity, where a user connects to a macro base station (MBS) and a small base station (SBS) for its DL and UL delivery respectively, flexible user association and high energy efficiency can be achieved.

\subsection{RAN Reconfiguration}
Upon using RAN decoupling technologies, conventional RANs have been softwarized and split into independent virtual NFs.
Then, to support diverse application (APPs), only the necessary decoupled NFs will be chosen and assembled to construct RANs on a demand basis.
For reconfiguring 6G RAN efficiently, there have been at least two possible solutions, namely RAN slicing and MANO.

RAN slicing \cite{4a} allows the virtual NFs and radio resources in RAN to be dynamically allocated to various services.
Explicitly, when a specific service request arrives, RAN slicing selects the necessary NFs using the resources allocated, and then encapsulates them to form a customized virtual network.

MANO is also an effective technique of achieving RAN configuration, and it can be applied more flexibly to different network scopes than RAN slicing.
Among them, the management part is responsible for the overall monitoring of RANs, including life cycle management, fault detection, accounting and security monitoring \cite{6a}.
The orchestration part \cite{27} supports the NFs and uses the underlying resources for reassembling the RAN, which mainly includes two sub-parts: choosing the appropriate NFs and appropriately scheduling them for a certain service.

\subsection{RAN Capability Enhancement}
MEC supports local computing and storage resources, which can be integrated with the open RAN both for providing the users with low-latency services and for reducing the transport network's burden \cite{20}.
Although the OS RAN architecture has not as yet been defined, an OS-MEC scheme will be established and detailed in our paper, which relies on the above decoupling and reconfiguration technologies.

Furthermore, 6G will have an ever increasing complexity in support of its compelling applications, heterogeneous networking architecture, and diverse resources. Hence the conventional network management methods (\textit{i.e.}, deployment, resource allocation and operation) associated with human-controlled philosophies will become inadequate, calling for sophisticated artificial intelligence (AI) techniques \cite{2} in the RAN, for achieving self-organizing, automated operation and CapEx/OpEx savings.

\section{The Open-Source MEC}
In this section, we incorporate the fundamental concepts of OS networks into MEC by relying on decoupling and reconfiguration, and propose the OS-MEC concept for improving the openness and flexibility of the emerging MEC systems.
Explicitly, by developing a service-based MEC layer, we decompose the tightly coupled service functions into multiple independent NFs and support them flexibly using a simple service-based interface (SBI).
Next, by introducing the convenient template and instance concept, NF's MANO is proposed for reconfiguring MEC, in which the OS-MEC's templates are predefined and then instantiated when needed.

\subsection{OS-MEC Framework}
\begin{figure}[!tp]
\setlength{\abovecaptionskip}{0cm}
\setlength{\belowcaptionskip}{0cm}
  \centering
  \includegraphics[width=3.53in]{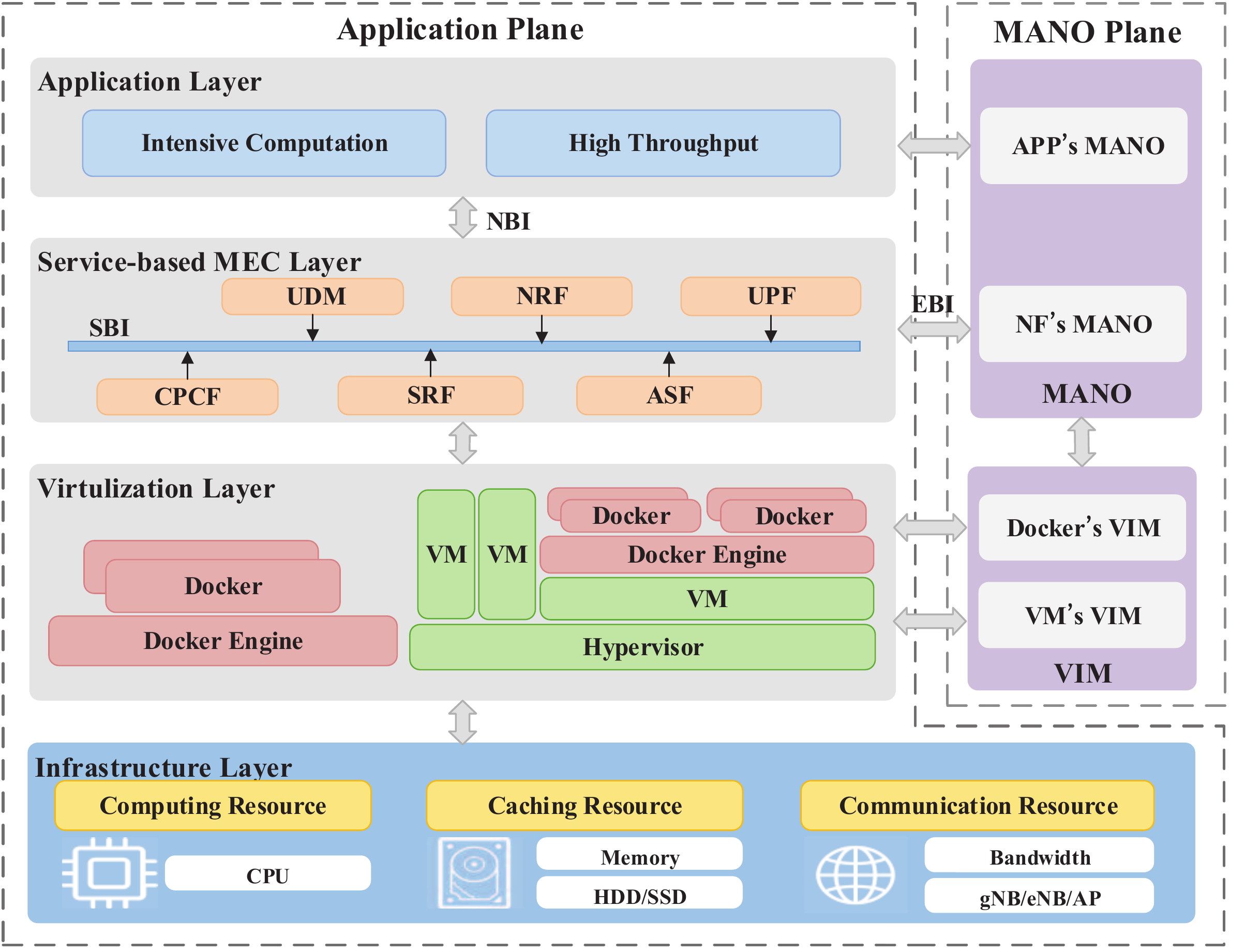}\\
  \caption{OS-MEC framework.}\label{fig2}
\end{figure}

The OS-MEC framework conceived is shown in Fig.~\ref{fig2}.
The entire framework includes the application plane and MANO plane, where the former is mainly responsible for data processing and transmission, involving the infrastructure layer, virtualization layer, service-based MEC layer, and application layer from bottom to top.
By contrast, the latter is composed of the virtualized infrastructure manager (VIM) and MANO, which completes the coordination and resource management of the OS-MEC system.
The pair of adjacent layers communicate with each other through standard interfaces and cooperate for completing the MEC services requested by users.

As seen in Fig.~\ref{fig2}, the infrastructure layer is at the lowest level, covering all computing, caching and communication resources of the system.
Specifically, the central processing unit (CPU) provides high-performance computing power for the service-based MEC layer right below the application layer at the top of Fig.~\ref{fig2}.
The memory and the hard disk drive (HDD)/solid state drive (SSD) are the main components of the caching resources in the infrastructure layer of Fig.~\ref{fig2}.
The communications resources of this layer include the bandwidth as well as the eNB, next-generation node base station (gNB) and access point (AP), as shown in the bottom layer of Fig.~\ref{fig2}.

At the virtualization layer of Fig.~\ref{fig2}, based on the NFV concept, the underlying three-dimensional resources may be decoupled from the dedicated hardware and abstracted into a resource pool that could be shared by various NFs at the service-based MEC layer of Fig.~\ref{fig2}.
Then, OS-MEC creates multiple Docker containers or virtual machines (VMs), which are run on the underlying resource pool and could simultaneously support different NFs for providing customized services \cite{8}, as seen at the virtualization layer of Fig.~\ref{fig2}.

The service-based MEC layer of Fig.~\ref{fig2} constitutes the core part of OS-MEC,
which includes a unified SBI and diverse NFs,
where the SBI can connect these NFs together based on the unified stateless hypertext transfer protocol (HTTP) for ensuring that they can communicate directly with each other when needed.
Furthermore, we decouple the originally centralized service functions into independent NFs.
Specifically, the user plane function (UPF) of the service-based MEC layer in Fig.~\ref{fig2} is borrowed from 5GC for OS-MEC, in order to lend the 5G new radio (NR) capabilities to OS-MEC.
Subsequently, to enhance OS-MEC, we download several other NFs at the control plane of 5GC and develop several new NFs.
These NFs are transparent to each other and can also be combined at will, so that they could be activated and released by all users instantly in support of their customized services.

Actually, the user does not have to rely on the specific details of OS-MEC.
Instead, they will use various APPs at the application layer to gain access to customized services.
For simplicity, only two APPs are considered in this paper for processing caching and computing requests, respectively, \textit{i.e.}, the high throughput and the intensive computation scenarios of the application layer seen at the top of Fig.~\ref{fig2}.

Additionally, the MANO plane at the right of Fig.~\ref{fig2} is responsible for managing these NFs as well as APPs, and for scheduling the VIM to allocate resources.
As mentioned earlier, MANO is in charge of MEC reconfiguration, where the OS-MEC's templates and instances are proposed for implementing flexible reconfiguration, as and when required.
The VIM of the MANO plane in Fig.~\ref{fig2} manages the virtualized resources according to MANO's commands, so as to ensure that appropriate computing, caching and communication resources are supplied for the upper layers of Fig.~\ref{fig2}.

Since there is rich literature on the design and implementation of both the virtualization layer and on the infrastructure layer \cite{1b,6d,11,12,15,18,19}, we mainly consider the service-based MEC layer and the MANO plane of OS-MEC.

\subsection{MEC Decoupling}
\begin{figure}[!tp]
\setlength{\abovecaptionskip}{0cm}
\setlength{\belowcaptionskip}{0cm}
  \centering
  \includegraphics[width=3.53in]{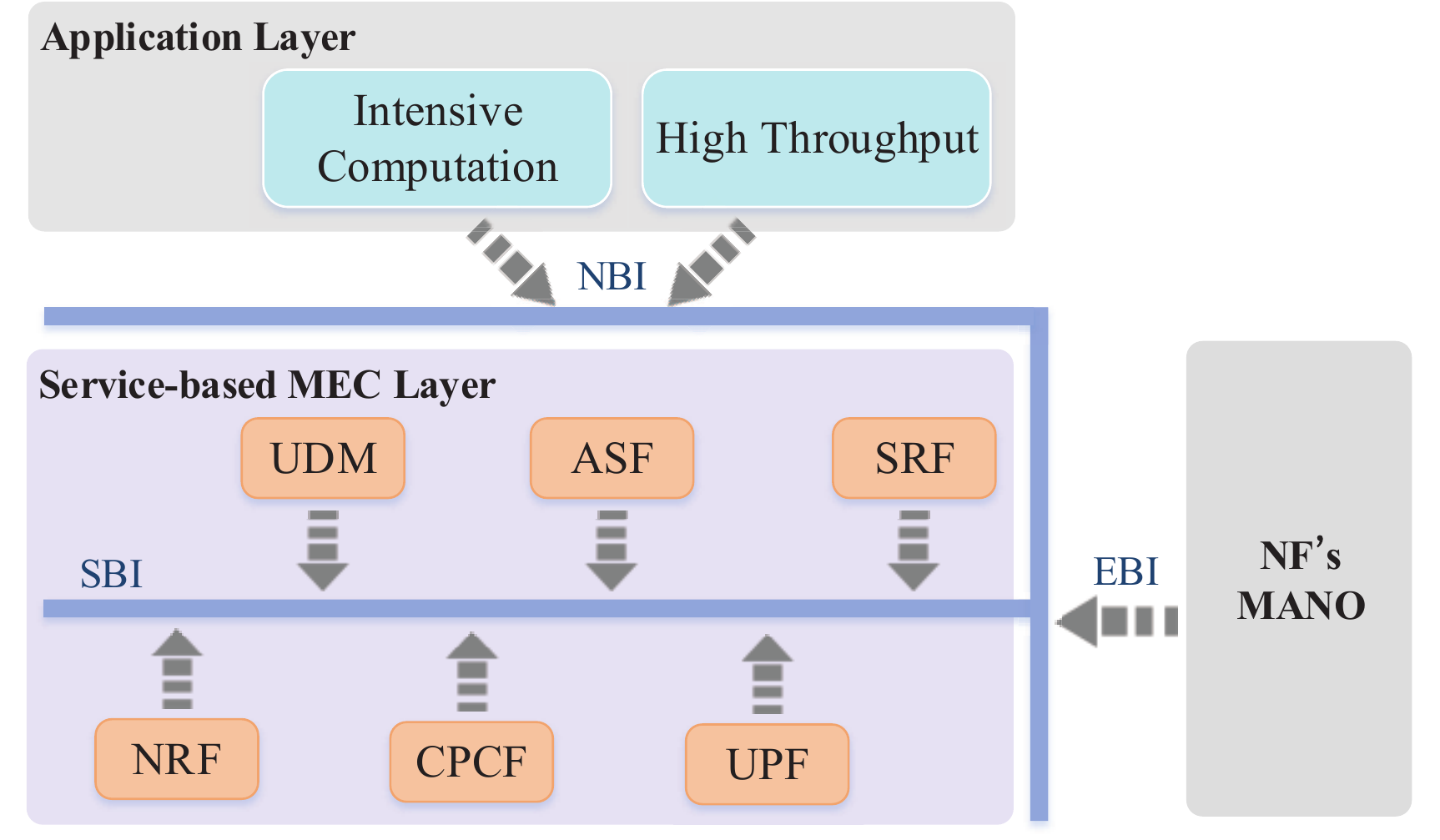}\\
  \caption{A detailed view of the service-based MEC layer of Fig.~\ref{fig2}.}\label{fig3}
\end{figure}

In order to decouple the MEC functions, we conceive the service-based MEC layer of the OS-MEC framework, which evolved from the traditional MEC layer, mainly composed of the SBI and the service-oriented NFs.
We extract the service-based MEC layer from Fig.~\ref{fig2}, and portray its detailed structure in Fig.~\ref{fig3}.

The SBI connects all the NFs together for facilitating their interface programming and implementation.
The north-bound interface (NBI) is an application program interface (API) provided for the application layer to interact with the service-based MEC Layer in order to obtain services.
The east-bound interface (EBI) is an API for NF's MANO to interact with the service-based MEC Layer, which is mainly used to manage and orchestrate both the resource allocation and life cycle of NFs.
On the basis of using the same HTTP, we design a unified Representational State Transfer-ful (RESTful) \cite{16} API for the SBI, NBI and EBI, which is lightweight and can be read both by people and machines easily.
Thus, the service-based MEC layer provides direct communications between the NFs, the MANO and the NFs, as well as the NFs and APPs,
thereby reducing the interface-complexity and the protocol-inconsistency in the traditional MEC architecture \cite{6}.

In this paper, all service-oriented NFs are independently deployed in the Docker containers mentioned in Fig.~\ref{fig2} in order to become virtualized.
There are six basic NFs in our proposed OS-MEC of Fig.~\ref{fig3}.
Specifically, the MEC NFs learned from 5GC include the unified data management (UDM), the NF repository function (NRF) and UPF seen in Fig.~\ref{fig3}.
Furthermore, the communication protocol conversion function (CPCF), the service registry function (SRF) and the application selection function (ASF) of Fig.~\ref{fig3} are newly developed after considering the context of OS-MEC.
Below we briefly describe the functional differences of UDM and NRF applied both in OS-MEC as well as 5GC, followed by the portrayal of our proposed CPCF, SRF and ASF of Fig.~\ref{fig3}.

\begin{figure*}[!tp]
\setlength{\abovecaptionskip}{0cm}
\setlength{\belowcaptionskip}{0cm}
  \centering
  \includegraphics[width=7.2in]{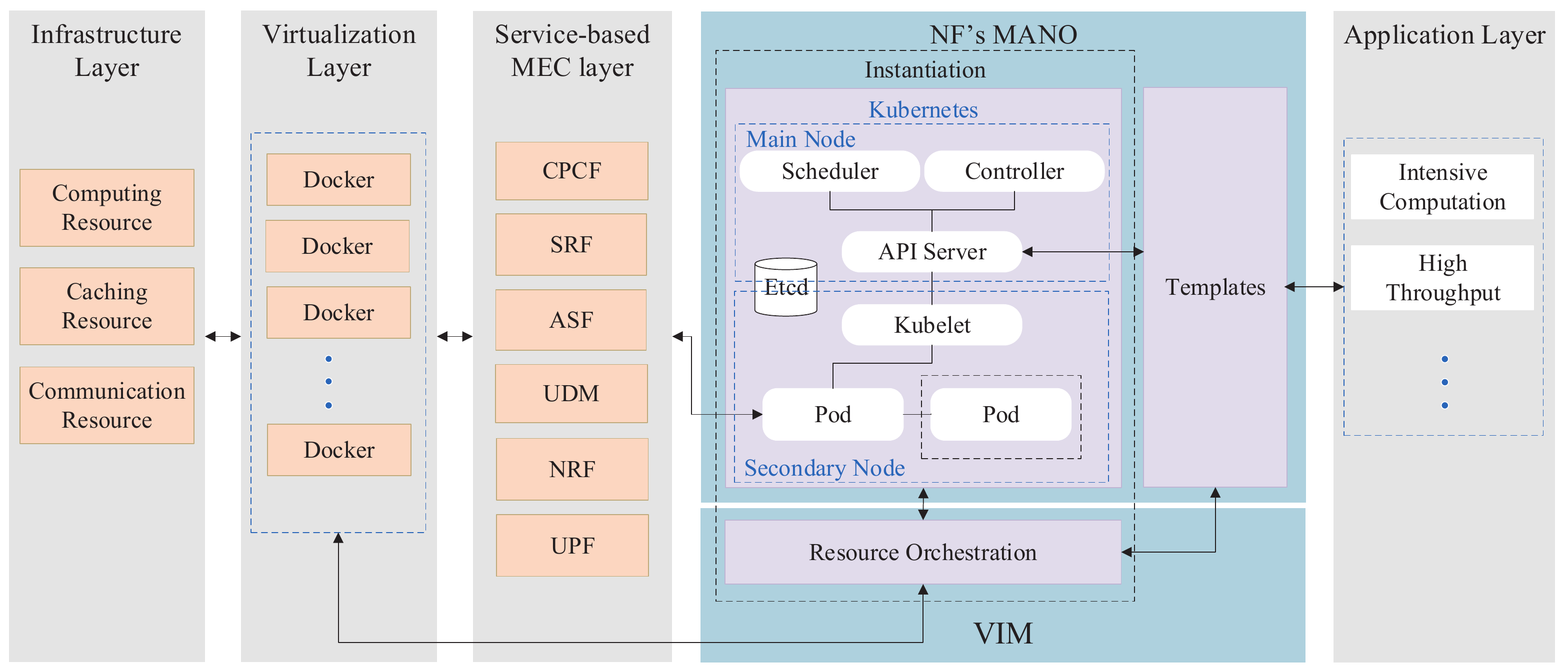}\\
  \caption{A detailed view of the NF's MANO of Fig.~\ref{fig2}.}\label{fig4}
\end{figure*}
\begin{itemize}
  \item UDM is used for unified data storage and management.
  In contrast to the 5GC UDM, the OS-MEC UDM does not have to deal with the issue of data structure compatibility because of its semi-homogeneous data.
  It mainly manages the data through the MySQL database management system \cite{8a}, and implements data insertion, search, update and deletion in the data tables of specific services. OS-MEC UDM can also provide standardized interfaces for SRF and ASF to ensure flexible access and real-time data update.
  \item NRF is a repository provided for centralized storing of NFs.
  OS-MEC further decouples the 5GC NRF, thus the OS-MEC NRF is only responsible for the storage of NFs and APPs, while the service registration is repackaged as SRF, and the service discovery is integrated into MANO.
  Due to the limited edge resources, general NFs (including SRF, CPCF, UDM and NRF) use local storage, while dedicated NFs, \textit{e.g.}, ASF and APPs are stored in the remote image repository.
  As for the latter, the OS-MEC server will rely on specific NF and APP images to provide services only when needed.
  \item CPCF can complete the communication protocol conversion. In the OS-MEC system, the stateless HTTP is mainly used for data exchange between NFs. However, the protocols of users requesting edge services are heterogeneous, so CPCF has to analyze and re-encapsulate them into a unified HTTP protocol for supporting compatibility of the system.
  \item SRF is responsible for the service registration of new APPs. When a new APP gets to OS-MEC, SRF first registers the APP to UDM and updates the APP-related parameters to data tables for improving the information exchange between APP and NFs. Then, SRF stores the APP in the remote NRF image repository and provides an external access interface for allowing all the other edge servers to use the image.
  \item ASF directly interacts with APPs through NBI for service selection. Specifically, compelling services could be provided in the form of APPs, such as intensive computation and high throughput services.
   Acting as the middleware between MEC APPs and other NFs, ASF is used both for the specific location of the requesting APPs and for communication between APPs and the other basic NFs.
\end{itemize}

The NF's MANO of Fig.~\ref{fig3} can flexibly schedule the NFs according to the specific service types, since the unified HTTP is used as EBI between the service-based MEC layer and NF's MANO, and is also used as SBI in the service-based MEC layer.
With all the NFs deployed in the Docker containers in our solution, Kubernetes \cite{9}, the open-source implementation of the container cluster management, is advocated for centrally managing all the Docker containers playing the role of MANO.
Its specific architecture will be introduced in the next subsection.

Based on the above NFs, we are ready for characterizing the basic functionalities of OS-MEC. Hence anyone could develop his/her own NFs for supporting new applications.

\subsection{MEC Reconfiguration}
As the MEC management is not directly related to the reconfiguration of OS-MEC, the joint management of MEC and NFV proposed by ETSI can still be used in OS-MEC,
albeit the conception of service-based MEC layers is bound to impose challenges on the orchestration scheme.
Therefore, by means of Kubernetes, we present OS-MEC's templates and instances of NF's MANO,
which supports the flexible scheduling of NFs, and any further MEC reconfiguration.
More explicitly, some templates are pre-defined for the corresponding applications, as exemplified by a computation-intensive template.
However, when no user asks for this application, we do not allocate the related resources to the template, hence, the template is empty.
By contrast, as shown in Fig.~\ref{fig4}, if a user requests this application, we shall activate the template by assigning computing, caching and communication resources as an instantiation.
\subsubsection{\textbf{NF's MANO}}
A detailed view of the NF's MANO extracted from Fig.~\ref{fig2} is shown in the middle (blue) part of Fig.~\ref{fig4},
where the right side represents the predefinition and selection of templates, while the left side represents the templates' instantiation based upon Kubernetes \cite{9}, which is a container-based cluster management platform.
The VIM seen at the bottom right of Fig.~\ref{fig4} is used for scheduling the underlying resource pool to arrange for the immediate resource deployment for instantiation, so as to avoid the assignment of excessive resources.

Observe in Fig.~\ref{fig4} that there is always a main node and either one or several secondary nodes in Kubernetes, where the former is in charge of managing Kubernetes, while the latter is in charge of the orchestration of NFs.
In a secondary node, there are two basic components dedicated to instantiation in NF's MANO, namely Pod and Kubelet as seen in Fig.~\ref{fig4}. As defined in \cite{9}, Pod is the smallest unit of  deploying and monitoring NFs in Kubernetes, which can run several Docker containers simultaneously. Kubelet \cite{9} of Fig.~\ref{fig4} maintains the life cycle of Pods and the communication between the master and the secondary nodes. In the master node, there are three basic components, namely the scheduler, controller and API server. The API server provides the only entry for secondary nodes to the application layer, \textit{i.e.}, Pods communicate with the given templates selected by the user through the API server for implementing MEC reconfiguration. The Scheduler and Controller of Fig.~\ref{fig4} manage and monitor all secondary nodes, \textit{e.g.}, for fault detection. Moreover, there is a Kubernetes' data center represented by Etcd of Fig.~\ref{fig4} for storing the states of the above components.

\subsubsection{\textbf{OS-MEC Template}}
\begin{figure}[!tp]
\setlength{\abovecaptionskip}{0cm}
\setlength{\belowcaptionskip}{0cm}
  \centering
  \includegraphics[width=3.53in]{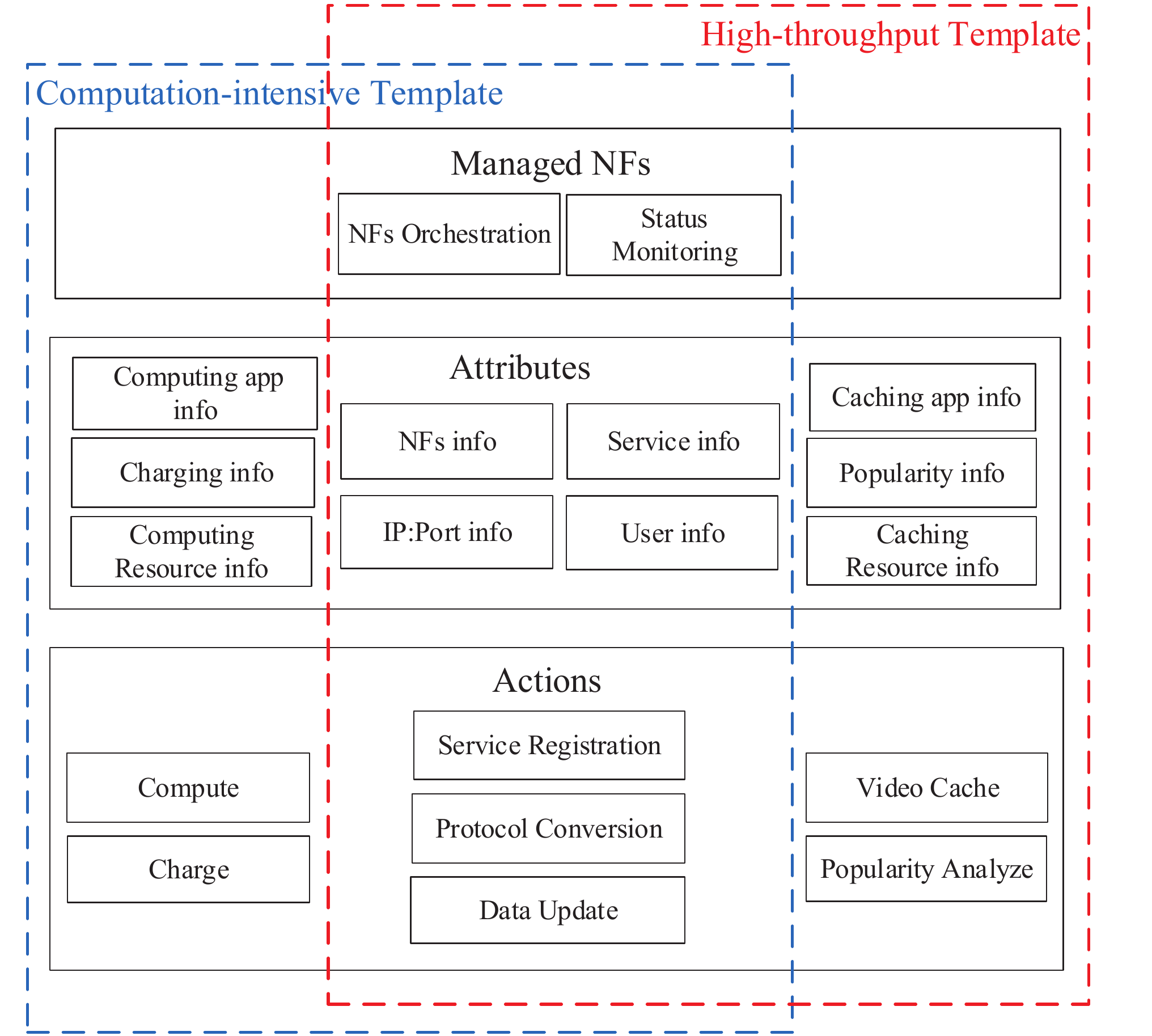}\\
  \caption{A detailed view of the OS-MEC templates introduced at the right of Fig.~\ref{fig4}.}\label{fig5}
\end{figure}

The template seen close to the right of Fig.~\ref{fig4} is provided for extracting the common characteristics of a class of problems.
Given this idea, a static OS-MEC template is predefined, \textit{e.g.}, for selecting appropriate NFs and for predefining the related parameters according to the specific type of APPs.
Again, we consider two APPs, namely high throughput and intensive computation, as seen at the right of Fig.~\ref{fig4}. These are examples of describing the OS-MEC templates, where the former is provided for high-throughput multimedia video services, while the latter is mainly for computing intensive services.
Naturally, there are some differences between the two templates in terms of the NFs selection and data table definition in UDM, as detailed in Fig.~\ref{fig5}.

Explicitly, observe in Fig.~\ref{fig5} that the OS-MEC template is composed of three tiers, including Managed NFs, Attributes and Actions.
In this context, all NFs have to communicate directly with the Managed NFs at the top layer, which are responsible for monitoring the templates' status and orchestrating NFs.
The middle-layer of Fig.~\ref{fig5} representing the Attributes acts as the database of templates, where the associated parameters are stored in the UDM's data tables.
Finally, the underlying Actions seen at the bottom of Fig.~\ref{fig5} are provided for completing a range of customized functions by selecting appropriate NFs as well as APPs and for updating the predefined UDM parameters in real time.

We partition the computation-intensive and the high-throughput templates into the shared parts (overlap of the templates in Fig.~\ref{fig5}) and the dedicated non-overlapping parts of each.
The shared part is irrelevant for the APPs,
it essentially ensures the normal operation of the system,
while the dedicated template is the key for supporting OS-MEC in providing specific services for the corresponding APP.
For example, both the computation-intensive and the high-throughput templates have to complete service registration provided by SRF, protocol conversion completed by CPCF and data updates by UDM.
However, the dedicated ASF and APP of the computation-intensive template are responsible for completing the computations based on the source-data, as seen at the bottom left of Fig.~\ref{fig5} and for setting up a charging function.
By contrast, the dedicated ASF and APP of the high-throughput template process video caching and analyze the video popularity, as observed at the bottom right of Fig.~\ref{fig5}.
\subsubsection{\textbf{OS-MEC Instance}}
The OS-MEC template of Fig.~\ref{fig5} cannot directly provide services and we do not allocate the related resources for the template.
As a result, users can only gain access to the edge services through instances.
The instantiation determines whether the OS-MEC system can run stably, which is the key for reconfiguring MEC.

The instantiation process mainly includes the following three steps.
Firstly, when a user requests edge services, such as intensive computation and high throughput, the MANO of Fig.~\ref{fig4} configures the operating environment of NFs, \textit{i.e.} the API Server of Fig.~\ref{fig4} locates the idle Pod and updates the relevant information to Etcd of Fig.~\ref{fig4}, which provides basic environmental support for instantiation.
Secondly, the MANO of Fig.~\ref{fig4} schedules VIM for allocating the specified virtual computing, caching and communication resources to users.
Finally, under the specifically configured operating environment and the resources allocated, MANO actually runs the predefined NFs and APPs of the template, \textit{i.e.}, Kubelet runs the Pod that specifically deploys NFs and APPs.
Therefore, the dedicated MEC instance has been created.
\begin{figure*}[!tp]
\setlength{\abovecaptionskip}{0cm}
\setlength{\belowcaptionskip}{0cm}
  \centering
  \includegraphics[width=6.5in]{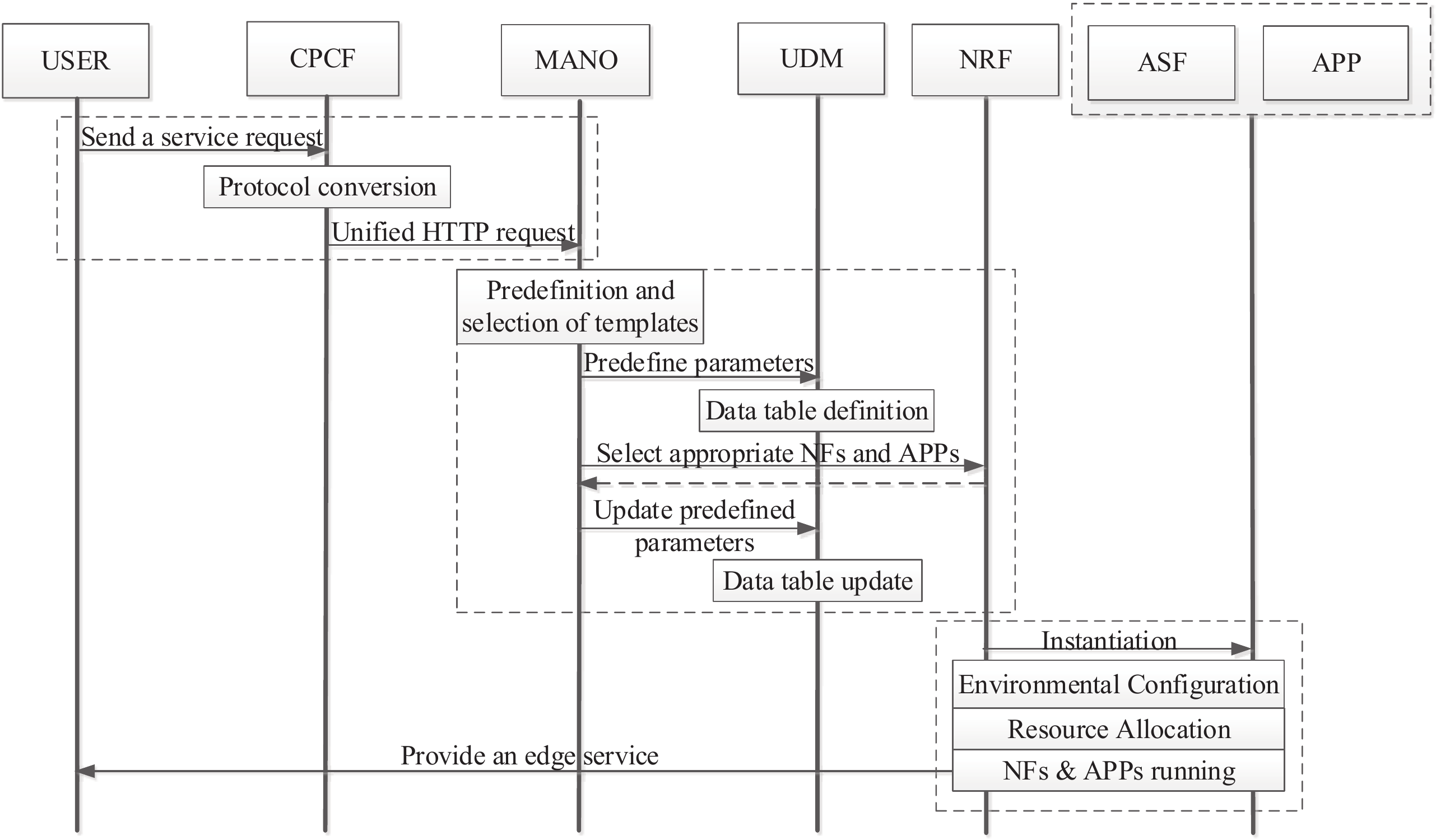}\\
  \caption{OS-MEC workflow.}\label{fig5_2}
\end{figure*}

In Fig.~\ref{fig5_2}, we summarize the workflow of OS-MEC, commencing from the instant when a user sends a service request until the edge service is instantiated.
Specifically,
\begin{enumerate}
\item when a user sends a service request to the OS-MEC, such as the intensive computation service request, CPCF firstly actions the protocol identification.
\item Upon receiving a HTTP request, CPCF does not need to perform the protocol conversion; otherwise, it re-encapsulates the service request using HTTP.
In this way, CPCF can send the unified HTTP request to MANO.
\item Then, MANO selects the computation-intensive template for the user. Explicitly,
    \begin{enumerate}
    \item the predefined parameters of this template are added to the specific data table in UDM.
    \item By querying the NFs and APP information specified by the template in UDM, MANO selects the appropriate NFs and APPs stored in NRF.
    \item MANO updates the predefined parameters in UDM's data table, completing the selection of templates.
    \end{enumerate}
\item Nextly, the instantiation process is implemented by the following three steps: the environmental configuration, resource allocation, and NF activation and APP activation.
\item Finally, OS-MEC provides a computation-intensive edge service to the user in the form of APP by the service interface.
\end{enumerate}
At this point, we have succeeded in creating the OS-MEC scheme by completing the MEC decoupling and reconfiguration.

\section{Demonstration in Test Network}

\begin{figure}[!tp]
\setlength{\abovecaptionskip}{0cm}
\setlength{\belowcaptionskip}{0cm}
  \centering
  \includegraphics[width=3in]{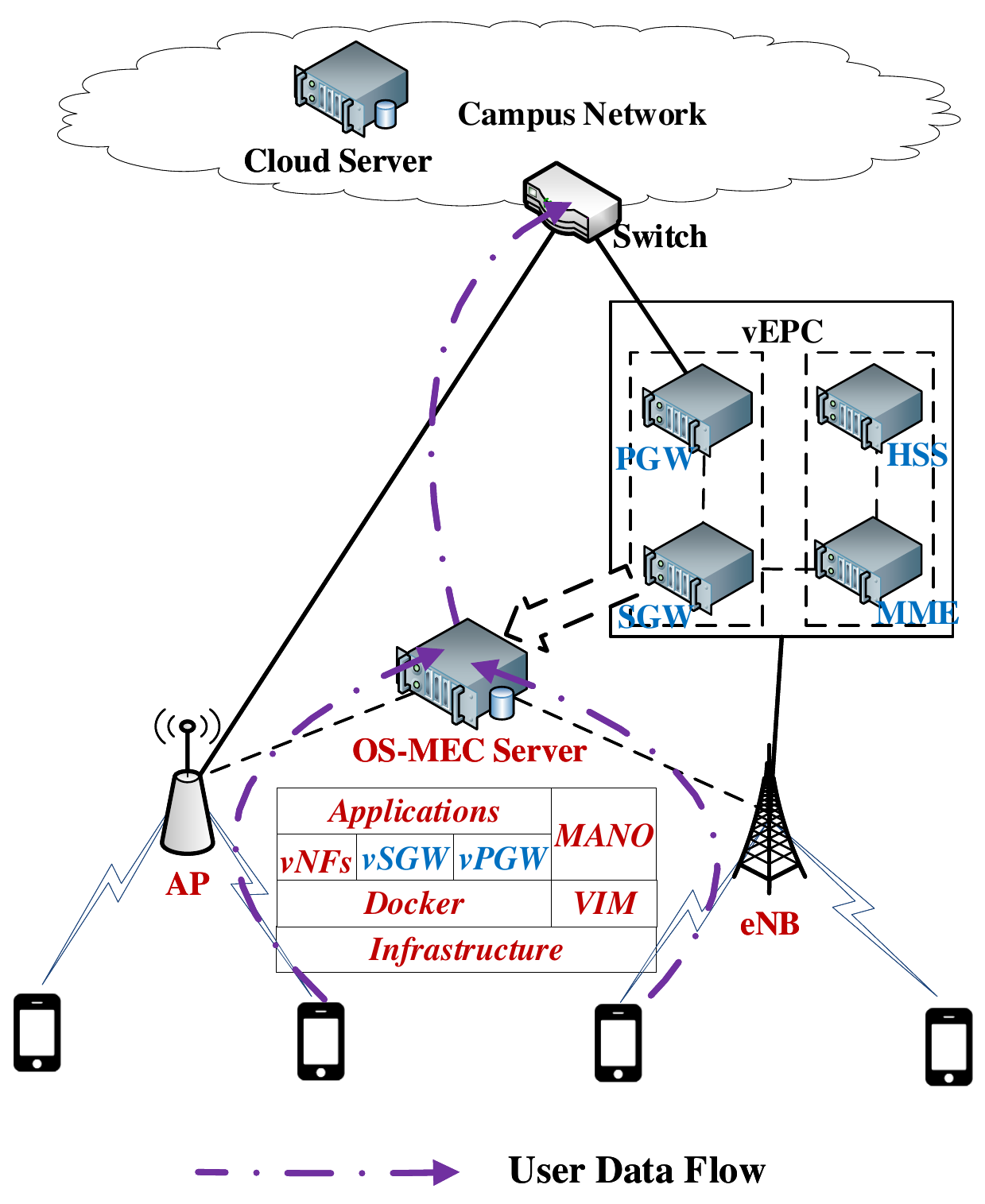}\\
  \caption{Test network.}\label{fig6}
\end{figure}

In this section, we critically appraise the proposed OS-MEC scheme in the test network deployed at Xidian University and then characterize a pair of use cases, namely the intensive computation and the high throughput scenarios.
Our test network is shown in Fig.~\ref{fig6}.
A 3GPP R10 based cellular network including an evolved packet core (EPC) and six eNBs was built based on open-source software provided by OAI \cite{10} and general-purpose hardware, namely an Intel Core i7-7700@3.6GHz CPU and 16 GB random access memory (RAM).
A commercial wireless router was used as an AP to deploy a WiFi network, which can support multiple access.
Then, we deploy all the network elements in Docker containers to arrange for them to become virtualized, yielding a virtualized EPC (vEPC) and virtual NFs (vNFs).
Subsequently, the user plane of both the serving gateway (SGW) and of the packet data network gateway (PGW) is moved from vEPC into the OS-MEC server using an Intel Core i7-7567U@3.5GHz CPU and 32 GB RAM.
The control plane, \textit{i.e.} the home subscriber server (HSS) and the mobility management entity (MME), remains in the vEPC.
Finally, based on Kubernetes, NF's MANO including its templates and instances is implemented to furnish users with customized services.

For instance, in the high throughput use case providing a video caching service, the user firstly accesses the test network of Fig.~\ref{fig6} by selecting an appropriate access mode, \textit{e.g.}, eNB.
Then, the network has to confirm, whether the requested video is stored in the OS-MEC server.
If so, the high throughput service may indeed be directly completed in the OS-MEC server by the specific workflow seen in Fig.~\ref{fig5_2}.
By contrast, if the requested video cannot be found in the OS-MEC server, then the vSGW and vPGW in the OS-MEC server will connect to the cloud server of the campus network through the switch to access the required resources.

In the intensive computation use case, we provide three services, \textit{i.e.} sum and prime sum computation as well as face recognition.
These three services are requested in three independent instances respectively, which are created by the same computation-intensive template though.
Therefore, three containers have to be started for the above instances, respectively.
By contrast, in the high throughput use case providing a video caching service, a user watches his/her video cached at the OS-MEC server, which only creates a single instance and runs a single container.

\begin{figure}[!tp]
\setlength{\abovecaptionskip}{0cm}
\setlength{\belowcaptionskip}{0cm}
  \centering
  \includegraphics[width=3.6in]{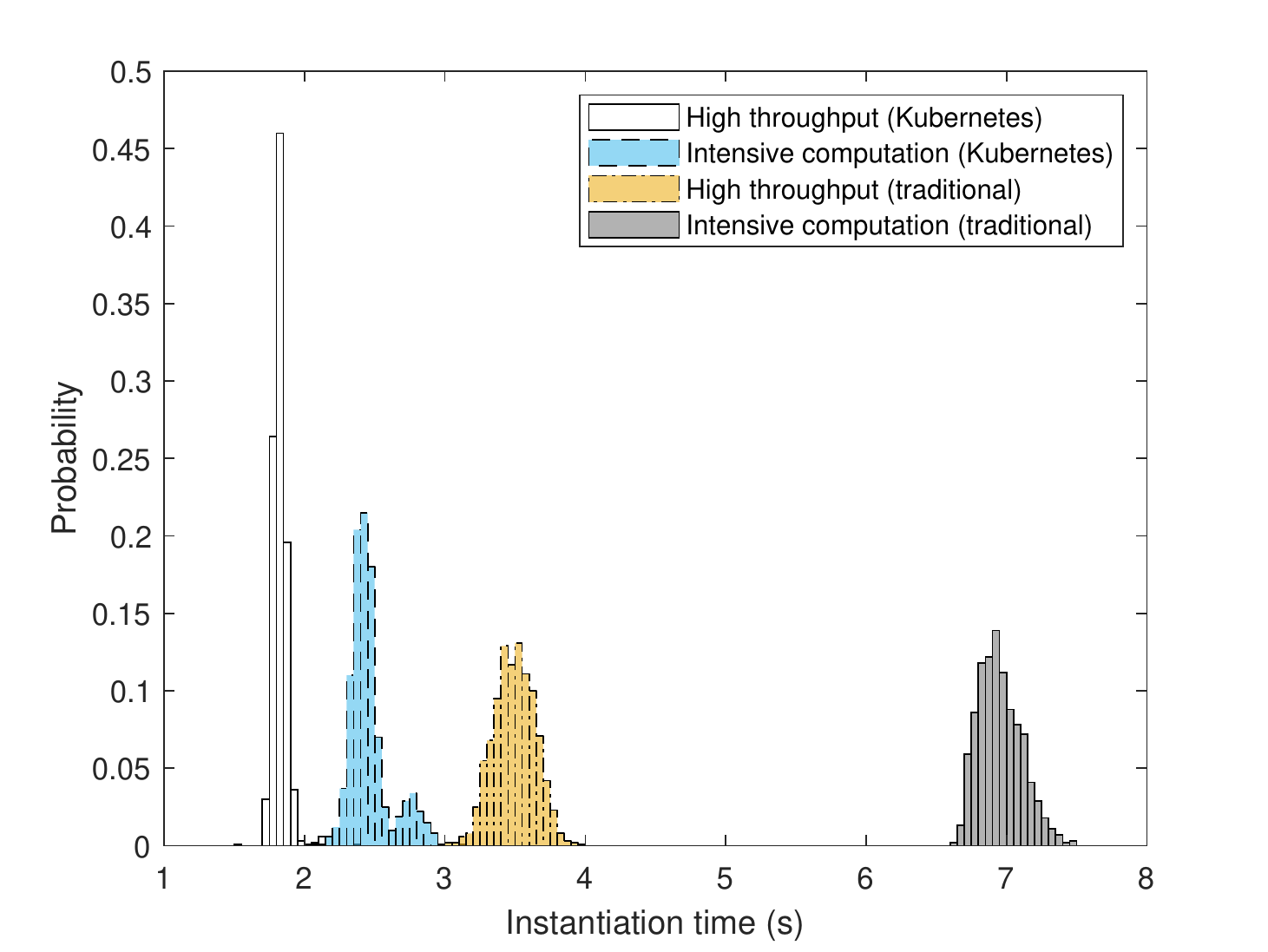}\\
  \caption{Instantiation time-duration for computation-intensive and high-throughput services.}\label{fig7_1}
\end{figure}
\begin{figure}[!tp]
\setlength{\abovecaptionskip}{0cm}
\setlength{\belowcaptionskip}{0cm}
  \centering
  \includegraphics[width=3.6in]{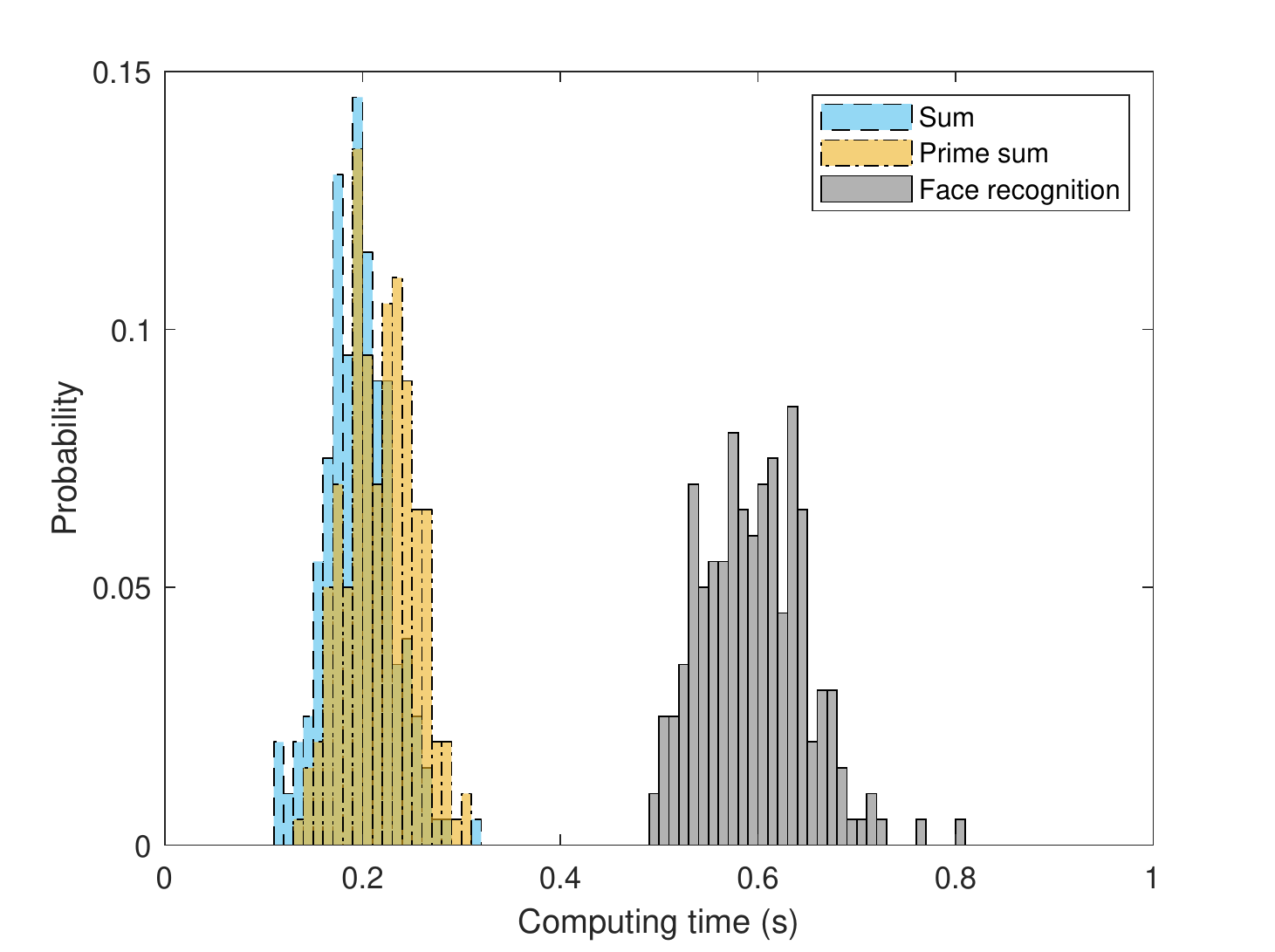}\\
  \caption{Computing time of three computation-intensive services.}\label{fig7_2}
\end{figure}
Fig.~\ref{fig7_1} shows a histogram for 1000 experiments for characterizing the resultant instantiation time-duration both in the proposed Kubernetes method and in the traditional script running method used in \cite{10},
where we simultaneously support three computation-intensive services to test the instantiation performance.
For both applications, the average instantiation duration of Kubernetes is always lower than that of the traditional script,
because the Kubernetes method can run multiple containers in parallel, while the traditional method only runs them separately through scripts.
Moreover, the intensive computation APP takes a little longer to instantiate, because it starts three containers at the same time, while the high throughput APP only runs a single container.
However, when using our Kubernetes method, their difference in time duration is small.
This allows the operators to easily incorporate new NFs or services as and when needed without degrading the user experience.

To provide further insights, Fig.~\ref{fig7_2} portrays the comparison of the computing time of three computation-intensive services by a histogram for 1000 experiments.
Since each picture has to be transmitted from mobile phones to the OS-MEC server and processed by the file transfer protocol (FTP), the average computing time of face recognition is higher than that of the other two services.

\begin{figure}[!tp]
\setlength{\abovecaptionskip}{0cm}
\setlength{\belowcaptionskip}{0cm}
  \centering
  \includegraphics[width=2.7in]{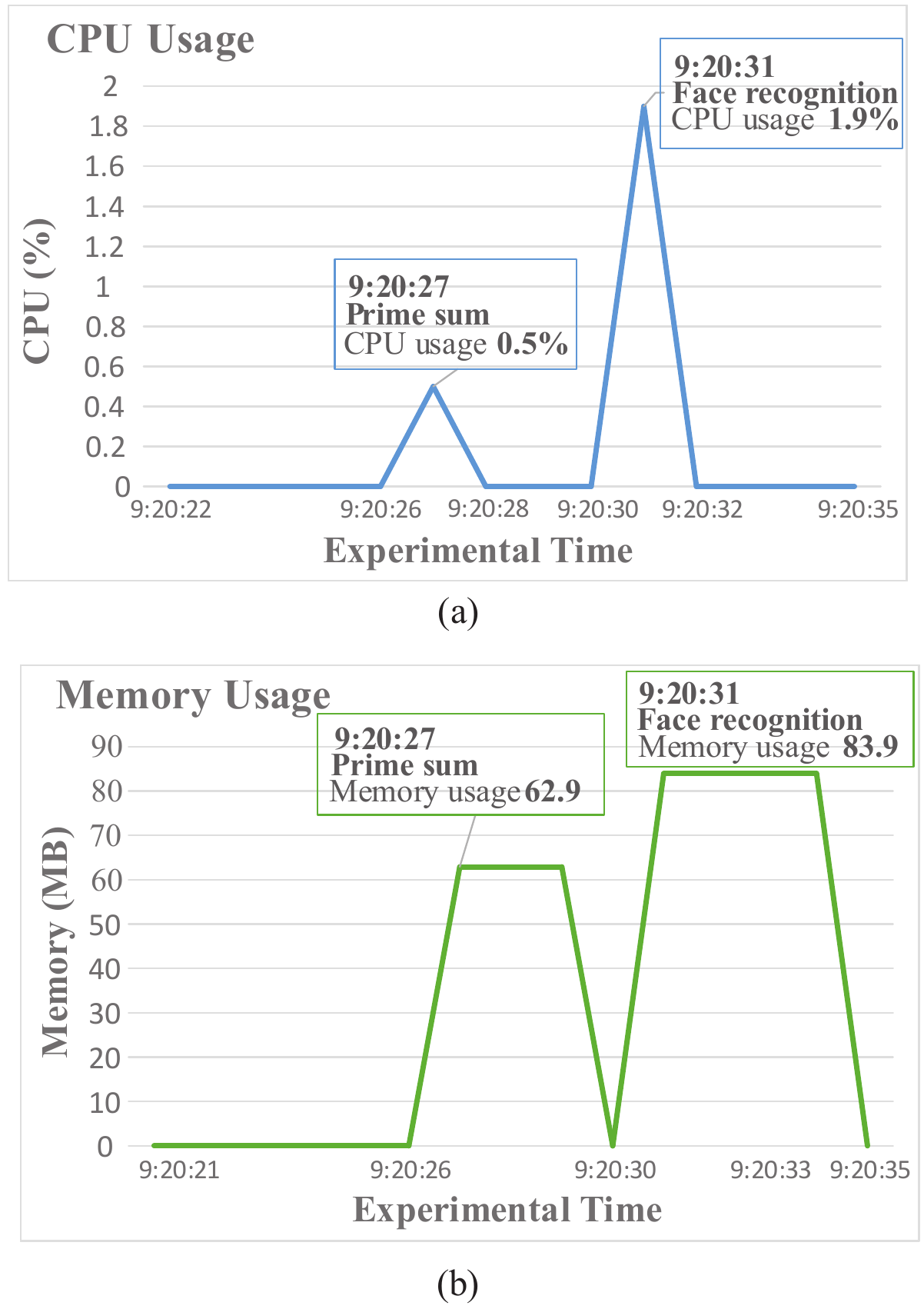}\\
  \caption{(a) CPU usage for intensive computation; (b) memory usage for intensive computation.}\label{fig8}
\end{figure}
Fig.~\ref{fig8} analyzes the CPU and memory usage by calculating the prime sum and face recognition in turn.
Since much more data have to be processed during face recognition, the CPU load of face recognition is about four times higher than that of prime sum computations.
The face recognition also has to store more data than the prime sum computation, so it requires more memory in Fig.~\ref{fig8}(b) (83.9MB).
Then, whenever the network completes data processing, the CPU is freed up immediately and the computing resources are released.
By contrast, the memory will not be freed up unless we handle it manually.
For example, as seen in Fig.~\ref{fig8}(a), the CPU is released immediately once the prime sum service is completed at 9:20:28.
However, in Fig.~\ref{fig8}(b), the memory remained reserved, until we released it manually at 9:20:30.
Therefore, the limited edge computing and caching resources of OS-MEC are used flexibly and efficiently.
\begin{figure}[!tp]
\setlength{\abovecaptionskip}{0cm}
\setlength{\belowcaptionskip}{0cm}
  \centering
  \includegraphics[width=3in]{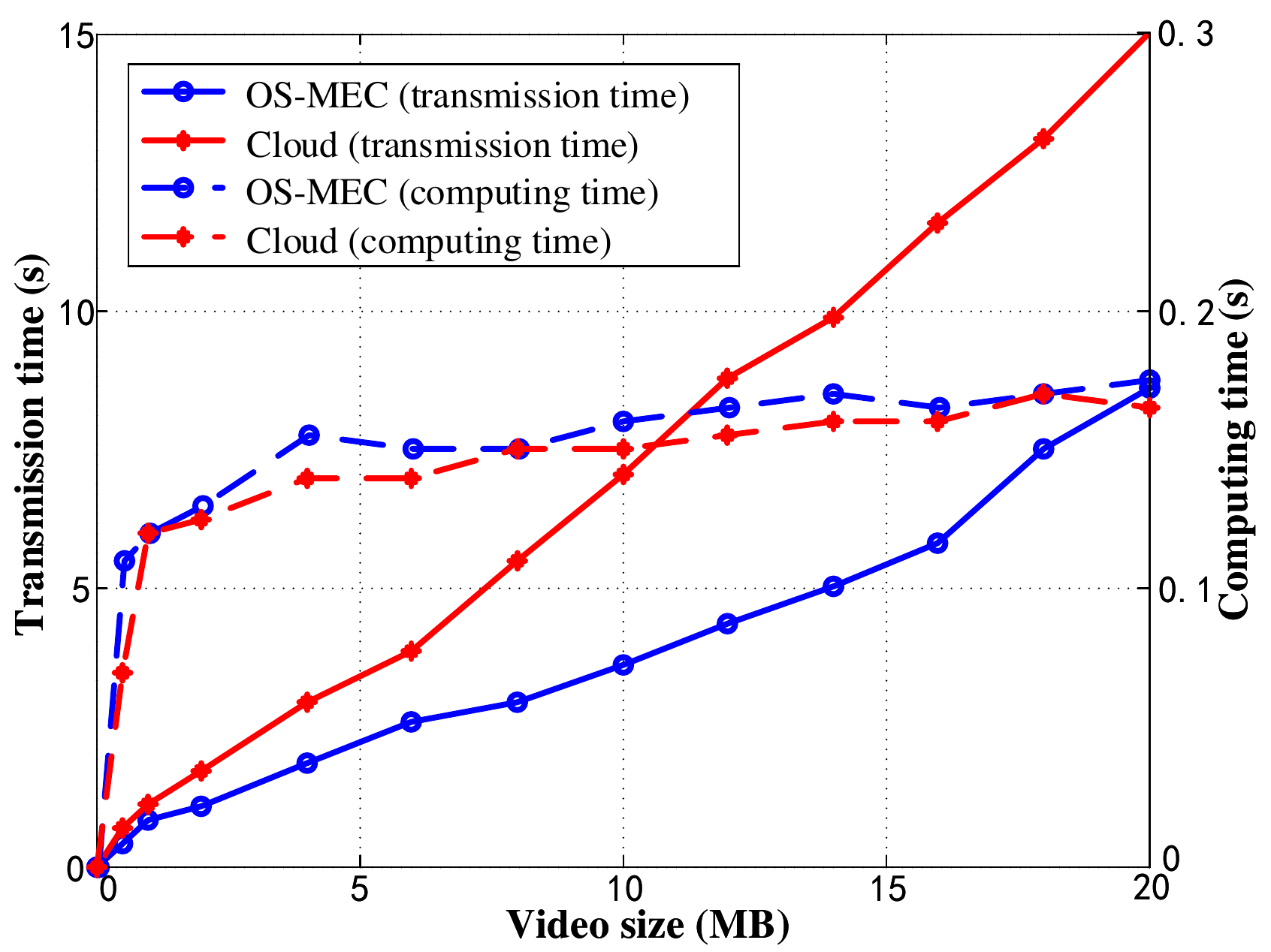}\\
  \caption{Transmission and computing time for high throughput.}\label{fig9}
\end{figure}

Fig.~\ref{fig9} quantifies the transmission and computing time of the high-throughput service, in which different-length videos are cached at the edge or cloud.
Upon increasing the video-length, we can see that the gap of the transmission time between the cloud and the OS-MEC scheme gradually increases, which shows that the advantage of our proposed scheme is gradually enhanced.
Since the computation loads of the high-throughput service are limited, the computing time of both the cloud and of the edge is only about 0.14s on average, and their difference is small.
Furthermore, the increase of video size has little effect on the computing time, which is basically in a stable state.

\vspace{-0in}
\section{Technical challenges}
As mentioned above, the rapidly evolving scenarios and requirements of the 6G network provides a compelling opportunity for introducing the OS cellular network concept into MEC, which allows the operators to adaptively customize the users' networks at the edge based on the proposed OS-MEC philosophy.
Our solutions have the potential of influencing the standardization of the MEC architecture for next-generation networks.
However, there are still some open technical challenges in the research of 6G-oriented OS cellular networks having a MEC functionality, which need further research and global discussion.

\subsection{Cloud-network-edge-terminal collaborative OS cellular network}
In recent years, substantial progress has been made in the research of OS networks.
For example, due to the proposal of the 5GC SBA, the OS CN architecture has been studied quite widely; the proposed OS-MEC scheme provides a complete overview of the edge-based OS architecture and its implementation.
Nevertheless, the OS RAN architecture has not been clearly defined due to its complex air interfaces, heterogeneous cell structure and incompatible network protocols.
Moreover, an integrated OS cellular network paradigm involving collaborations between the cloud, the network, the edge and the terminal is inevitable in the 6G network of the near future.
Whether the existing OS network components can be integrated and work together still needs further testing.

\subsection{Effective integration with 6G scenarios}
The 6G system will further improve the existing 5G networking scenarios \cite{24,25},
with the objective of further improving ultra-reliable low latency communications (URLLC).
It will also embrace new paradigm shifts, \textit{e.g.} space-air-ground integrated networks \cite{42}, terahertz (THz) communication \cite{43}, \textit{etc}.
This requires that the future 6G OS network must be sufficiently flexible and intelligent for seamlessly supporting new scenarios and components, or for upgrading the existing functions without causing global disruption to the architecture.
However, the proposed OS-MEC had to be developed and operated on the existing general-purpose hardware and open-source software platforms.
Therefore, its integration with the future 6G application scenarios and paradigms needs further research.

\subsection{Open-source market fragmentation}
The research of OS network platforms has attracted the attention of numerous organizations and institutions.
Numerous OS network platforms have been developed, such as O-RAN, ONAP, OPNFV and the proposed OS-MEC, \textit{etc}.
However, the availability of a bewildering gamut of platforms can lead to the fragmentation of the open-source market, hence resulting in interoperability problems.
In order to circumvent these problems, on the one hand, a unified open-source standard should be established for supporting openness and interoperability.
On the other hand, it requires closer cooperation among open-source organizations to coordinate all the OS network platforms in the near future.

\subsection{Cybersecurity and intellectual property protection}
The 6G-oriented OS cellular network architecture is expected to integrate all operating networks and provide users with customized capabilities including mobile services, critical industrial infrastructure, vertical sectors and other new network services.
However, these new network features and the wide application of 6G are fraught with potential security and privacy challenges.
In addition, the OS architecture means that based on general-purpose hardware, any potential user is able to study, update and release new software versions to other users in a collaborative open-access manner.
This however highlights the virtualization security issues, and poses a challenge in terms of the definition and protection of intellectual property.
Therefore, how to effectively protect cybersecurity and intellectual property is still a huge challenge.

\subsection{Engaging Telecom Vendors}
Although more and more researchers and practitioners rely on open-source components to build customized cellular networks supporting MEC functionality, telecom vendors are not overly motivated to follow this approach.
Since this open-source mode makes the specialized hardware associated with particular operating systems or functions in the original traditional network obsolete, vendors can no longer make profits by developing different specialized hardware.
Therefore, how to persuade telecom vendors to join this open-source movement from an economic perspective and to further expand the target group supporting this movement is still a challenge.

\vspace{-0in}
\section{Conclusions and Future Research}
In this paper, we have conceived the OS-MEC scheme for 6G relying on sophisticated decoupling and reconfiguration principles.
First of all, an OS-MEC framework was presented based on ETSI's MEC architecture.
Then, by means of the SBA and NFV technology, the tightly coupled service functions were decomposed into independent NFs.
Next, by introducing the template and instantiation based concepts into MANO, the disaggregated NFs may be re-assembled and the required resources may be allocated in order to provide customized services for users.
In this context, the evolved NFs were illustrated in the service-based MEC layer.
The key components and the associated implementation processes were described in the NF's MANO.
Then, a pair of typical use cases have been critically appraised through demonstrations in a small-scale test network.
Finally, we have discussed some potential technical challenges of 6G-oriented OS cellular networks.

In our future work, we will aim for developing the OS RAN architecture in order to construct a complete OS cellular network for 6G.
Based on the twinned core principles of the OS RAN architecture: RAN decoupling and RAN reconfiguration, we envisage to propose a new three-dimensional RAN slicing mechanism, which realizes the decoupling and re-encapsulation of resources and NFs from the three dimensions of resource virtualization, function virtualization and network slicing, so as to provide customized virtual RANs for users.
We hope to fill the gaps in the research of OS RAN architecture and lay the foundations for the ultimate realization of cloud-network-edge-terminal collaborative OS cellular networks.

\section*{List of Abbreviations}
   \vspace{0.1cm}
         \noindent %
        \begin{tabular}{>{\raggedright}p{2cm}>{\raggedright}p{6.2cm}}
            3GPP & 3rd Generation Partnership Project \tabularnewline
        \end{tabular}
        \noindent %
        \begin{tabular}{>{\raggedright}p{2cm}>{\raggedright}p{6.2cm}}
            4G & Fourth Generation \tabularnewline
        \end{tabular}
        \noindent %
        \begin{tabular}{>{\raggedright}p{2cm}>{\raggedright}p{6.2cm}}
            5G & Fifth Generation \tabularnewline
        \end{tabular}
        \noindent %
        \begin{tabular}{>{\raggedright}p{2cm}>{\raggedright}p{6.2cm}}
            5GC & 5G Core Network \tabularnewline
        \end{tabular}
        \noindent %
        \begin{tabular}{>{\raggedright}p{2cm}>{\raggedright}p{6.2cm}}
            6G & Sixth Generation \tabularnewline
        \end{tabular}
         \noindent %
        \begin{tabular}{>{\raggedright}p{2cm}>{\raggedright}p{6.2cm}}
            AP & Access Point \tabularnewline
        \end{tabular}
        \noindent %
        \begin{tabular}{>{\raggedright}p{2cm}>{\raggedright}p{6.2cm}}
            API & Application Program Interface \tabularnewline
        \end{tabular}
         \noindent %
         \begin{tabular}{>{\raggedright}p{2cm}>{\raggedright}p{6.2cm}}
             APP & Application \tabularnewline
         \end{tabular}
         \noindent %
        \begin{tabular}{>{\raggedright}p{2cm}>{\raggedright}p{6.2cm}}
            ASF & Application Selection Function \tabularnewline
        \end{tabular}
         \noindent %
        \begin{tabular}{>{\raggedright}p{2cm}>{\raggedright}p{6.2cm}}
            CBS & Control Base Station \tabularnewline
        \end{tabular}
         \noindent %
        \begin{tabular}{>{\raggedright}p{2cm}>{\raggedright}p{6.2cm}}
            CDUS & Central Unit and Distributed Unit Splitting \tabularnewline
        \end{tabular}
         \noindent %
        \begin{tabular}{>{\raggedright}p{2cm}>{\raggedright}p{6.2cm}}
            CN & Core Network \tabularnewline
        \end{tabular}
         \noindent %
        \begin{tabular}{>{\raggedright}p{2cm}>{\raggedright}p{6.2cm}}
            COTS & Commercial Off-The-Shelf \tabularnewline
        \end{tabular}
        \noindent %
        \begin{tabular}{>{\raggedright}p{2cm}>{\raggedright}p{6.2cm}}
            CP & Control Plane \tabularnewline
        \end{tabular}
        \noindent %
        \begin{tabular}{>{\raggedright}p{2cm}>{\raggedright}p{6.2cm}}
            CPCF & Communication Protocol Conversion Function \tabularnewline
        \end{tabular}
        \noindent %
        \begin{tabular}{>{\raggedright}p{2cm}>{\raggedright}p{6.2cm}}
            CPU & Central Processing Unit \tabularnewline
        \end{tabular}
        \noindent %
        \begin{tabular}{>{\raggedright}p{2cm}>{\raggedright}p{6.2cm}}
            CUPS & Control Plane and User Plane Splitting \tabularnewline
        \end{tabular}
        \noindent %
        \begin{tabular}{>{\raggedright}p{2cm}>{\raggedright}p{6.2cm}}
            DBS & Data Base Station \tabularnewline
        \end{tabular}
        \noindent %
        \begin{tabular}{>{\raggedright}p{2cm}>{\raggedright}p{6.2cm}}
            DL & Downlink \tabularnewline
        \end{tabular}
        \noindent %
        \begin{tabular}{>{\raggedright}p{2cm}>{\raggedright}p{6.2cm}}
            DUDe & Downlink and Uplink Decoupling \tabularnewline
        \end{tabular}
         \noindent %
        \begin{tabular}{>{\raggedright}p{2cm}>{\raggedright}p{6.2cm}}
            EBI & East-Bound Interface \tabularnewline
        \end{tabular}
        \noindent %
        \begin{tabular}{>{\raggedright}p{2cm}>{\raggedright}p{6.2cm}}
            eNB &  evolved Node Base Station \tabularnewline
        \end{tabular}
        \noindent %
        \begin{tabular}{>{\raggedright}p{2cm}>{\raggedright}p{6.2cm}}
            EPC & Evolved Packet Core \tabularnewline
        \end{tabular}
        \noindent %
        \begin{tabular}{>{\raggedright}p{2cm}>{\raggedright}p{6.2cm}}
            ETSI & European Telecommunications Standards Institute \tabularnewline
        \end{tabular}
        \noindent %
        \begin{tabular}{>{\raggedright}p{2cm}>{\raggedright}p{6.2cm}}
            FTP & File Transfer Protocol \tabularnewline
        \end{tabular}
        \noindent %
        \begin{tabular}{>{\raggedright}p{2cm}>{\raggedright}p{6.2cm}}
            gNB & Next-generation Node Base Station \tabularnewline
        \end{tabular}
        \noindent %
        \begin{tabular}{>{\raggedright}p{2cm}>{\raggedright}p{6.2cm}}
            HDD & Hard Disk Drive \tabularnewline
        \end{tabular}
        \noindent %
        \begin{tabular}{>{\raggedright}p{2cm}>{\raggedright}p{6.2cm}}
            HSDe &  Hardware/Software Decoupling \tabularnewline
        \end{tabular}
        \noindent %
        \begin{tabular}{>{\raggedright}p{2cm}>{\raggedright}p{6.2cm}}
            HSS & Home Subscriber Server \tabularnewline
        \end{tabular}
        \noindent %
        \begin{tabular}{>{\raggedright}p{2cm}>{\raggedright}p{6.2cm}}
            HTTP & Hypertext Transfer Protocol \tabularnewline
        \end{tabular}
        \noindent %
        \begin{tabular}{>{\raggedright}p{2cm}>{\raggedright}p{6.2cm}}
            LF & Linux Foundation \tabularnewline
        \end{tabular}
        \noindent %
        \begin{tabular}{>{\raggedright}p{2cm}>{\raggedright}p{6.2cm}}
            MANO & Management and Orchestration \tabularnewline
        \end{tabular}
        \noindent %
        \begin{tabular}{>{\raggedright}p{2cm}>{\raggedright}p{6.2cm}}
            MBS & Macro Base Station \tabularnewline
        \end{tabular}
   	    \noindent
   	      \begin{tabular}{>{\raggedright}p{2cm}>{\raggedright}p{6.2cm}}
            MEC & Multi-access Edge Computing \tabularnewline
        \end{tabular}
        \noindent %
        \begin{tabular}{>{\raggedright}p{2cm}>{\raggedright}p{6.2cm}}
            MME & Mobility Management Entity \tabularnewline
        \end{tabular}
        \noindent %
        \begin{tabular}{>{\raggedright}p{2cm}>{\raggedright}p{6.2cm}}
            NBI & North-Bound Interface \tabularnewline
        \end{tabular}
        \noindent %
        \begin{tabular}{>{\raggedright}p{2cm}>{\raggedright}p{6.2cm}}
            NDN-ECC & Named Data Networking-based Edge Cloud Computing \tabularnewline
        \end{tabular}
        \noindent %
        \begin{tabular}{>{\raggedright}p{2cm}>{\raggedright}p{6.2cm}}
            NF & Network Function \tabularnewline
        \end{tabular}
        \noindent %
        \begin{tabular}{>{\raggedright}p{2cm}>{\raggedright}p{6.2cm}}
            NFV & Network Function Virtualization\tabularnewline
        \end{tabular}
        \noindent %
        \begin{tabular}{>{\raggedright}p{2cm}>{\raggedright}p{6.2cm}}
            NR & New Radio \tabularnewline
        \end{tabular}
        \noindent %
        \begin{tabular}{>{\raggedright}p{2cm}>{\raggedright}p{6.2cm}}
            NRF & NF Repository Function \tabularnewline
        \end{tabular}
        \noindent %
        \begin{tabular}{>{\raggedright}p{2cm}>{\raggedright}p{6.2cm}}
            OAI & Open-Air-Interface \tabularnewline
        \end{tabular}
        \noindent %
        \begin{tabular}{>{\raggedright}p{2cm}>{\raggedright}p{6.2cm}}
            ONAP & Open Network Automation Platform \tabularnewline
        \end{tabular}
        \noindent %
        \begin{tabular}{>{\raggedright}p{2cm}>{\raggedright}p{6.2cm}}
            OS-MEC & Open-source MEC \tabularnewline
        \end{tabular}
        \noindent %
        \begin{tabular}{>{\raggedright}p{2cm}>{\raggedright}p{6.2cm}}
            OPNFV & Open Platform of NFV \tabularnewline
        \end{tabular}
        \noindent %
        \begin{tabular}{>{\raggedright}p{2cm}>{\raggedright}p{6.2cm}}
            OS & Open-Source \tabularnewline
        \end{tabular}
        \noindent %
        \begin{tabular}{>{\raggedright}p{2cm}>{\raggedright}p{6.2cm}}
            PGW & Packet Data Network Gateway \tabularnewline
        \end{tabular}
        \noindent %
        \begin{tabular}{>{\raggedright}p{2cm}>{\raggedright}p{6.2cm}}
            RAM & Random Access Memory \tabularnewline
        \end{tabular}
        \noindent %
        \begin{tabular}{>{\raggedright}p{2cm}>{\raggedright}p{6.2cm}}
            RAN & Radio Access Networks \tabularnewline
        \end{tabular}
        \noindent %
        \begin{tabular}{>{\raggedright}p{2cm}>{\raggedright}p{6.2cm}}
            RESTful & Representational State Transfer-ful \tabularnewline
        \end{tabular}
        \noindent %
        \begin{tabular}{>{\raggedright}p{2cm}>{\raggedright}p{6.2cm}}
            SBA & Service-Based Architecture \tabularnewline
        \end{tabular}
        \noindent %
        \begin{tabular}{>{\raggedright}p{2cm}>{\raggedright}p{6.2cm}}
            SBI & Service-Based Interface \tabularnewline
        \end{tabular}
        \noindent %
        \begin{tabular}{>{\raggedright}p{2cm}>{\raggedright}p{6.2cm}}
            SBS & Small Base Station \tabularnewline
        \end{tabular}
        \noindent %
        \begin{tabular}{>{\raggedright}p{2cm}>{\raggedright}p{6.2cm}}
            SDEC & Software-Defined Edge Computing \tabularnewline
        \end{tabular}
        \noindent %
        \begin{tabular}{>{\raggedright}p{2cm}>{\raggedright}p{6.2cm}}
            SDN & Software-Defined Networking \tabularnewline
        \end{tabular}
        \noindent %
        \begin{tabular}{>{\raggedright}p{2cm}>{\raggedright}p{6.2cm}}
            SGW & Serving Gateway \tabularnewline
        \end{tabular}
        \noindent %
        \begin{tabular}{>{\raggedright}p{2cm}>{\raggedright}p{6.2cm}}
            SRF & Service Registry Function \tabularnewline
        \end{tabular}
        \noindent %
        \begin{tabular}{>{\raggedright}p{2cm}>{\raggedright}p{6.2cm}}
            SSD & Solid State Drive \tabularnewline
        \end{tabular}
        \noindent %
        \begin{tabular}{>{\raggedright}p{2cm}>{\raggedright}p{6.2cm}}
            THz & Terahertz \tabularnewline
        \end{tabular}
        \noindent %
        \begin{tabular}{>{\raggedright}p{2cm}>{\raggedright}p{6.2cm}}
            TN & Transport Network \tabularnewline
        \end{tabular}
        \noindent %
        \begin{tabular}{>{\raggedright}p{2cm}>{\raggedright}p{6.2cm}}
            UDM & Unified Data Management \tabularnewline
        \end{tabular}
        \noindent %
        \begin{tabular}{>{\raggedright}p{2cm}>{\raggedright}p{6.2cm}}
            UHD & Ultra High Definition \tabularnewline
        \end{tabular}
        \noindent %
        \begin{tabular}{>{\raggedright}p{2cm}>{\raggedright}p{6.2cm}}
            UL & Uplink \tabularnewline
        \end{tabular}
        \noindent %
        \begin{tabular}{>{\raggedright}p{2cm}>{\raggedright}p{6.2cm}}
            UP & User Plane \tabularnewline
        \end{tabular}
        \noindent %
        \begin{tabular}{>{\raggedright}p{2cm}>{\raggedright}p{6.2cm}}
            UPF & User Plane Function \tabularnewline
        \end{tabular}
        \noindent %
         \begin{tabular}{>{\raggedright}p{2cm}>{\raggedright}p{6.2cm}}
            URLLC & Ultra Reliable Low Latency Communications \tabularnewline
        \end{tabular}
        \noindent %
        \begin{tabular}{>{\raggedright}p{2cm}>{\raggedright}p{6.2cm}}
            vEPC & virtualized EPC \tabularnewline
        \end{tabular}
        \noindent %
        \begin{tabular}{>{\raggedright}p{2cm}>{\raggedright}p{6.2cm}}
            VIM & Virtualized Infrastructure Manager \tabularnewline
        \end{tabular}
        \noindent %
        \begin{tabular}{>{\raggedright}p{2cm}>{\raggedright}p{6.2cm}}
            VM & Virtual Machine \tabularnewline
        \end{tabular}
        \noindent %
        \begin{tabular}{>{\raggedright}p{2cm}>{\raggedright}p{6.2cm}}
            vNF & vitual NF \tabularnewline
        \end{tabular}
         \noindent %

\begin{IEEEbiography}[{\includegraphics[width=1in,height=1.25in,clip,keepaspectratio]{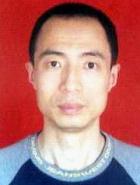}}]{\textbf{Liqiang Zhao}}
(Member, IEEE) received the B.Sc. degree in Electrical Engineering from Shanghai Jiaotong University, China, in 1992, the M.Sc. degree in Communications and Information Systems and the Ph.D. degree in Information and Communications Engineering from Xidian University, China, in 2000 and 2003, respectively. In 2008, he was awarded by the Program for New Century Excellent Talents in University, Ministry of Education, China.

From 1992 to 2005, he was a Research Engineer with the $20^{th}$ Research Institute, Chinese Electronics Technology Group Corporation (CETC), China. His research focused on mobile communication systems and spread spectrum communications. From 2005 to 2007, he was an Associate Professor with State Key Laboratory of Integrated Service Networks (ISN), Xidian University, China. His research focused on WiMAX, WLAN, and wireless sensor network. He was appointed as a Marie Curie Research Fellow at the Centre for Wireless Network Design (CWiND), University of Bedfordshire in June 2007 to conduct research in the GAWIND project funded under EU FP6 HRM programme. His activities focused on the area of automatic wireless broadband access network planning and optimization. Since June 2008, he has returned Xidian University, first as an Associate Professor and later as a Professor. His current research focuses on broadband wireless communications and space communications, \textit{etc}. He has more than 100 published in authorized academic periodicals both in and abroad and in international science conferences, wherein 30 of which are retrieved in SCI, and more than 70 of them are EI indexed, and 6 national invention patents. He has hosted / participated many national research projects, such as the National Natural Science Foundation, the 863 Program and the National Science Technology Major Projects, and several international research projects including the EU FP6, FP7 plans for international cooperation exchange projects, and some research projects from companies, \textit{e.g.}, Huawei.
\end{IEEEbiography}

\begin{IEEEbiography}[{\includegraphics[width=1in,height=1.25in,clip,keepaspectratio]{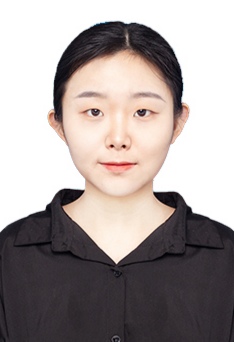}}]{\textbf{Guorong Zhou}}
is currently pursuing the Ph.D. degree from the school of Telecommunication Engineering, Xidian University, China.
Since April 2021, she has been a visiting Ph.D. student at the School of Electronics and Computer Science, University of Southampton, Southampton, U.K., on the funds from the Program of the China Scholarships Council.

Her research interests include wireless communications technologies, network slicing, multi-access edge computing and machine learning for communications.
\end{IEEEbiography}

\begin{IEEEbiography}[{\includegraphics[width=1in,height=1.25in,clip,keepaspectratio]{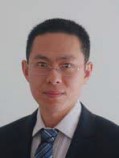}}]{\textbf{Gan Zheng}}
(Fellow, IEEE) received the BEng and the MEng from Tianjin University, Tianjin, China, in 2002 and 2004, respectively, both in Electronic and Information Engineering, and the PhD degree in Electrical and Electronic Engineering from The University of Hong Kong in 2008. He is currently Reader of Signal Processing for Wireless Communications in the Wolfson School of Mechanical, Electrical and Manufacturing Engineering, Loughborough University, UK. His research interests include machine learning for communications, UAV communications, mobile edge caching, full-duplex radio, and wireless power transfer. He is the first recipient for the 2013 IEEE Signal Processing Letters Best Paper Award, and he also received 2015 GLOBECOM Best Paper Award, and 2018 IEEE Technical Committee on Green Communications \& Computing Best Paper Award. He was listed as a Highly Cited Researcher by Thomson Reuters/Clarivate Analytics in 2019 and he is an IEEE Fellow. He currently serves as an Associate Editor for IEEE Communications Letters and IEEE Wireless Communications Letters.
\end{IEEEbiography}

\begin{IEEEbiography}[{\includegraphics[width=1in,height=1.25in,clip,keepaspectratio]{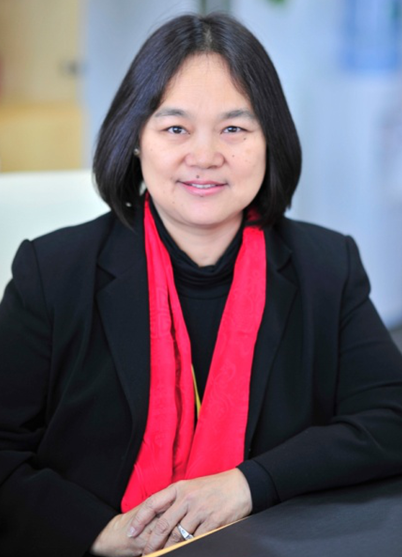}}]{\textbf{Chih-Lin I}}
(Fellow, IEEE) is CMCC Chief Scientist of Wireless Technologies. She received Ph.D. EE from Stanford University. She has won 2005 IEEE ComSoc Stephen Rice Prize, 2018 IEEE ComSoc Fred W. Ellersick Prize, the 7th IEEE Asia-Pacific Outstanding Paper Award, and 2015 IEEE Industrial Innovation Award for Leadership and Innovation in Next-Generation Cellular Wireless Networks.

She is the Chair of O-RAN Technical Steering Committee and an O-RAN Executive Committee Member, the Chair of FuTURE 5G/6G SIG, the Chair of WAIA (Wireless AI Alliance) Executive Committee, an Executive Board Member of GreenTouch, a Network Operator Council Founding Member of ETSI NFV, a Steering Board Member and Vice Chair of WWRF, a Steering Committee member and the Publication Chair of IEEE 5G and Future Networks Initiatives, the Founding Chair of IEEE WCNC Steering Committee, the Director of IEEE ComSoc Meetings and Conferences Board, a Senior Editor of IEEE Trans. Green Comm. \& Networking, a member of IEEE ComSoc SDB, SPC, and CSCN-SC, and a Scientific Advisory Board Member of Singapore NRF.

She has published over 200 papers in scientific journals, book chapters and conferences and holds over 100 patents. She is co-author of the book ``Green and Software-defined Wireless Networks -- From Theory to Practice'' and has also Co-edited two books: ``Ultra-dense Networks -- Principles and Applications'' and ``5G Networks -- Fundamental Requirements, Enabling Technologies, and Operations Management''. She is a Fellow of IEEE and a Fellow of WWRF. Her current research interests center around ICDT Deep Convergence: ``From Green \& Soft to Open \& Smart''.
\end{IEEEbiography}

\begin{IEEEbiography}[{\includegraphics[width=1in,height=1.25in,clip,keepaspectratio]{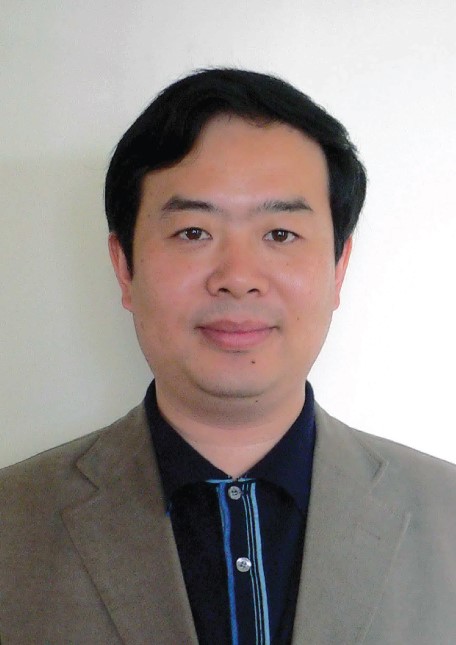}}]{\textbf{Xiaohu You}}
(Fellow, IEEE) received the B.S., M.S. and Ph.D. degrees in electrical engineering from Nanjing Institute of Technology, Nanjing, China, in 1982, 1985, and 1989, respectively. From 1987 to 1989, he was with Nanjing Institute of Technology as a Lecturer. From 1990 to the present time, he has been with Southeast University, first as an Associate Professor and later as a Professor. His research interests include mobile communications, adaptive signal processing, and artificial neural networks with applications to communications and biomedical engineering. He is the Chief of the Technical Group of China 3G/B3G Mobile Communication R \& D Project. He received the excellent paper prize from the China Institute of Communications in 1987 and the Elite Outstanding Young Teacher Awards from Southeast University in 1990, 1991, and 1993. He was also a recipient of the 1989 Young Teacher Award of Fok Ying Tung Education Foundation, State Education Commission of China.
\end{IEEEbiography}

\begin{IEEEbiography}[{\includegraphics[width=1in,height=1.25in,clip,keepaspectratio]{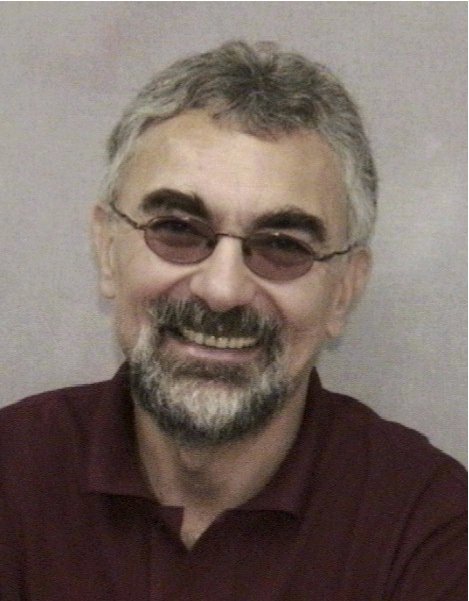}}]{\textbf{Lajos Hanzo}}
(Fellow, IEEE) received the master's and Ph.D. degrees from the Technical University (TU) of Budapest, in 1976 and 1983, respectively, and the D.Sc. degree from the University of Southampton, in 2004. He has published more than 1900 contributions at IEEE Xplore and 19 Wiley-IEEE Press books, and has helped the fast-track career of 123 Ph.D. students. Over 40 of them are professors at various stages of their careers in academia and many of them are leading scientists in the wireless industry. He is a Fellow of the Royal Academy of Engineering, IET, and EURASIP. He is a Foreign Member of the Hungarian Academy of Sciences and the former Editor-in-Chief of the IEEE Press. He has served several terms as the Governor of both IEEE ComSoc and VTS. He was awarded the Honorary Doctorates by the TU of Budapest, in 2009, and by the University of Edinburgh, in 2015.
\end{IEEEbiography}

\EOD


\vspace{-0in}
\begin{thebibliography}{1}
\bibitem{1a}
T. Taleb, K. Samdanis, B. Mada, H. Flinck, S. Dutta and D. Sabella, ``On Multi-Access Edge Computing: A Survey of the Emerging 5G Network Edge Cloud Architecture and Orchestration,'' \textit{IEEE Communications Surveys \& Tutorials}, vol. 19, no. 3, pp. 1657-1681, thirdquarter 2017.
\bibitem{111}
P. Porambage, J. Okwuibe, M. Liyanage, M. Ylianttila and T. Taleb, ``Survey on Multi-Access Edge Computing for Internet of Things Realization,'' \textit{IEEE Communications Surveys \& Tutorials}, vol. 20, no. 4, pp. 2961-2991, Fourthquarter 2018.
\bibitem{222}
H. A. Alameddine, S. Sharafeddine, S. Sebbah, S. Ayoubi and C. Assi, ``Dynamic Task Offloading and Scheduling for Low-Latency IoT Services in Multi-Access Edge Computing,'' \textit{IEEE Journal on Selected Areas in Communications}, vol. 37, no. 3, pp. 668-682, March 2019.
\bibitem{1b}
``Mobile edge computing: A key technology towards 5G,'' White Paper, ETSI, Sophia Antipolis, France, Sep. 2015.
\bibitem{21}
A. Ksentini and P. A. Frangoudis, ``On Extending ETSI MEC to Support LoRa for Efficient IoT Application Deployment at the Edge,'' \textit{IEEE Communications Standards Magazine}, vol. 4, no. 2, pp. 57-63, June 2020.
\bibitem{22}
G. Avino \textit{et al.}, ``A MEC-Based Extended Virtual Sensing for Automotive Services,'' \textit{IEEE Transactions on Network and Service Management}, vol. 16, no. 4, pp. 1450-1463, Dec. 2019.
\bibitem{23}
S. Yang, Y. Tseng, C. Huang and W. Lin, ``Multi-Access Edge Computing Enhanced Video Streaming: Proof-of-Concept Implementation and Prediction/QoE Models,'' \textit{IEEE Transactions on Vehicular Technology}, vol. 68, no. 2, pp. 1888-1902, Feb. 2019.
\bibitem{1}
A. Dogra, R. K. Jha and S. Jain, ``A Survey on Beyond 5G Network With the Advent of 6G: Architecture and Emerging Technologies,'' \textit{IEEE Access}, vol. 9, pp. 67512-67547, 2021.
\bibitem{2}
W. Saad, M. Bennis and M. Chen, ``A Vision of 6G Wireless Systems: Applications, Trends, Technologies, and Open Research Problems,'' \textit{IEEE Network}, vol. 34, no. 3, pp. 134-142, May/June 2020.
\bibitem{30}
T. S. Rappaport \textit{et al.}, ``Wireless Communications and Applications Above 100 GHz: Opportunities and Challenges for 6G and Beyond,'' \textit{IEEE Access}, vol. 7, pp. 78729-78757, 2019.
\bibitem{333}
Z. Lv and W. Xiu, ``Interaction of Edge-Cloud Computing Based on SDN and NFV for Next Generation IoT,'' \textit{IEEE Internet of Things Journal}, vol. 7, no. 7, pp. 5706-5712, July 2020.
\bibitem{24}
X. You, CX. Wang, J. Huang \textit{et al.}, ``Towards 6G wireless communication networks: vision, enabling technologies, and new paradigm shifts,'' \textit{Sci. China Inf. Sci.}, vol. 64, pp. 110301, Jan. 2021.
\bibitem{25}
C. Xu \textit{et al.}, ``Sixty Years of Coherent Versus Non-Coherent Tradeoffs and the Road From 5G to Wireless Futures,'' \textit{IEEE Access}, vol. 7, pp. 178246-178299, 2019.
\bibitem{26}
Stephanos Androutsellis-Theotoki; Diomides Spinellis; Maria Kechagia; Georgios Gousios, \textit{Open Source Software: A Survey from 10,000 Feet}, now, 2011.
\bibitem{3}
A. U. Rehman, R. L. Aguiar and J. P. Barraca, ``Network Functions Virtualization: The Long Road to Commercial Deployments,'' \textit{IEEE Access}, vol. 7, pp. 60439-60464, 2019.
\bibitem{4}
J. H. Cox \textit{et al.}, ``Advancing Software-Defined Networks: A Survey,'' \textit{IEEE Access}, vol. 5, pp. 25487-25526, 2017.
\bibitem{4a}
R. Su \textit{et al.}, ``Resource Allocation for Network Slicing in 5G Telecommunication Networks: A Survey of Principles and Models,'' \textit{IEEE Network}, vol. 33, no. 6, pp. 172-179, Nov.-Dec. 2019.
\bibitem{6}
B. Ali, M. A. Gregory and S. Li, ``Multi-Access Edge Computing Architecture, Data Security and Privacy: A Review,'' \textit{IEEE Access}, vol. 9, pp. 18706-18721, 2021.
\bibitem{7}
White Paper: 5G Americas' ``The Status of Open Source for 5G,'' Feb. 21, 2019.
\bibitem{7aa}
C. I, S. Kuklinsk$\acute{\textrm{i}}$, and T. Chen, ``A Perspective of O-RAN Integration with MEC, SON, and Network Slicing in the 5G Era,'' \textit{IEEE Network}, vol. 34, no. 6, pp. 3-4, Nov.-Dec. 2020.
\bibitem{6d}
Q. Pham \textit{et al.}, ``A Survey of Multi-Access Edge Computing in 5G and Beyond: Fundamentals, Technology Integration, and State-of-the-Art,'' \textit{IEEE Access}, vol. 8, pp. 116974-117017, 2020.
\bibitem{6c}
3GPP, ``System Architecture for the 5G system; Stage 2,'' TS 23.501, 2018.
\bibitem{16}
H. C. Rudolph, A. Kunz, L. L. Iacono and H. V. Nguyen, ``Security Challenges of the 3GPP 5G Service Based Architecture," \textit{IEEE Communications Standards Magazine}, vol. 3, no. 1, pp. 60-65, March 2019.

\bibitem{11}
W. Wang and H. Li, ``Light-Weight Platform for Attack Validation in LTE Network,'' \textit{IEEE Networking Letters}, vol. 2, no. 4, pp. 212-215, Dec. 2020.
\bibitem{12}
P. Kiri Taksande, P. Jha, A. Karandikar and P. Chaporkar, ``Open5G: A Software-Defined Networking Protocol for 5G Multi-RAT Wireless Networks,'' 2020 IEEE Wireless Communications and Networking Conference Workshops (WCNCW), 2020, pp. 1-6.
\bibitem{13}
S. Lag$\acute{\textrm{e}}$n, L. Giupponi, A. Hansson and X. Gelabert, ``Modulation Compression in Next Generation RAN: Air Interface and Fronthaul Trade-offs,'' \textit{IEEE Communications Magazine}, vol. 59, no. 1, pp. 89-95, January 2021.
\bibitem{14}
M. Pattaranantakul, R. He, Z. Zhang, A. Meddahi and P. Wang, ``Leveraging Network Functions Virtualization Orchestrators to Achieve Software-Defined Access Control in the Clouds,'' \textit{IEEE Transactions on Dependable and Secure Computing}, vol. 18, no. 1, pp. 372-383, 1 Jan.-Feb. 2021.
\bibitem{15}
A. Boubendir, E. Bertin and N. Simoni, ``On-demand dynamic network service deployment over NaaS architecture,'' NOMS 2016 - 2016 IEEE/IFIP Network Operations and Management Symposium, 2016, pp. 1023-1024.
\bibitem{18}
P. Hu, W. Chen, C. He, Y. Li and H. Ning, ``Software-Defined Edge Computing (SDEC): Principle, Open IoT System Architecture, Applications, and Challenges,'' \textit{IEEE Internet of Things Journal}, vol. 7, no. 7, pp. 5934-5945, July 2020.
\bibitem{19}
R. Ullah, M. A. U. Rehman and B. Kim, ``Design and Implementation of an Open Source Framework and Prototype For Named Data Networking-Based Edge Cloud Computing System,'' \textit{IEEE Access}, vol. 7, pp. 57741-57759, 2019.
\bibitem{5}
R. Mijumbi, J. Serrat, J. Gorricho, N. Bouten, F. De Turck and R. Boutaba, ``Network Function Virtualization: State-of-the-Art and Research Challenges,'' \textit{IEEE Communications Surveys \& Tutorials}, vol. 18, no. 1, pp. 236-262, Firstquarter 2016.
\bibitem{6b}
L. Yan, X. Fang, and Y. Fang, ``Stable Beamforming with Low Overhead for C/U-Plane Decoupled HSR Wireless Networks,'' \textit{IEEE Transactions on Vehicular Technology}, vol. 67, no. 7, pp. 6075-6086, July 2018.
\bibitem{5b}
H. Gupta, M. Sharma, A. Franklin A. and B. R. Tamma, ``Apt-RAN: A Flexible Split-Based 5G RAN to Minimize Energy Consumption and Handovers,'' \textit{IEEE Transactions on Network and Service Management}, vol. 17, no. 1, pp. 473-487, March 2020.
\bibitem{5c}
M. A. Lema, E. Pardo, O. Galinina, S. Andreev and M. Dohler, ``Flexible Dual-Connectivity Spectrum Aggregation for Decoupled Uplink and Downlink Access in 5G Heterogeneous Systems,'' \textit{IEEE Journal on Selected Areas in Communications}, vol. 34, no. 11, pp. 2851-2865, Nov. 2016.
\bibitem{6a}
A. Ksentini and P. A. Frangoudis, ``Toward Slicing-Enabled Multi-Access Edge Computing in 5G,'' \textit{IEEE Network}, vol. 34, no. 2, pp. 99-105, March/April 2020.
\bibitem{27}
A. J. Gonzalez, G. Nencioni, A. Kamisi$\acute{\textrm{n}}$ski, B. E. Helvik and P. E. Heegaard, ``Dependability of the NFV Orchestrator: State of the Art and Research Challenges,'' \textit{IEEE Communications Surveys \& Tutorials}, vol. 20, no. 4, pp. 3307-3329, Fourthquarter 2018.
\bibitem{20}
A. Ndikumana \textit{et al.}, ``Joint Communication, Computation, Caching, and Control in Big Data Multi-Access Edge Computing,'' \textit{IEEE Transactions on Mobile Computing}, vol. 19, no. 6, pp. 1359-1374, 1 June 2020.
\bibitem{8}
Y. Mao, C. You, J. Zhang, K. Huang and K. B. Letaief, ``A Survey on Mobile Edge Computing: The Communication Perspective,'' \textit{IEEE Communications Surveys \& Tutorials}, vol. 19, no. 4, pp. 2322-2358, Fourthquarter 2017.
\bibitem{8a}
S. Rautmare and D. M. Bhalerao, ``MySQL and NoSQL database comparison for IoT application,'' 2016 IEEE International Conference on Advances in Computer Applications (ICACA), 2016, pp. 235-238.
\bibitem{9}
K. Kaur, S. Garg, G. Kaddoum, S. H. Ahmed and M. Atiquzzaman, ``KEIDS: Kubernetes-Based Energy and Interference Driven Scheduler for Industrial IoT in Edge-Cloud Ecosystem,'' \textit{IEEE Internet of Things Journal}, vol. 7, no. 5, pp. 4228-4237, May 2020.
\bibitem{10}
T. Li, L. Zhao, R. Duan and H. Tian, ``SBA-Based Mobile Edge Computing,'' 2019 IEEE Globecom Workshops (GC Wkshps), Waikoloa, HI, USA, 2019, pp. 1-6.
\bibitem{42}
M. Giordani and M. Zorzi,``Non-Terrestrial Networks in the 6G Era: Challenges and Opportunities,'' \textit{IEEE Network}, vol. 35, no. 2, pp. 244-251, March/April 2021.
\bibitem{43}
Q. Xia, Z. Hossain, M. Medley and J. M. Jornet, ``A Link-Layer Synchronization and Medium Access Control Protocol for Terahertz-Band Communication Networks,'' \textit{IEEE Transactions on Mobile Computing}, vol. 20, no. 1, pp. 2-18, 1 Jan. 2021.


\end{thebibliography}
\end{document}